\documentclass[12pt]{iopart}

\usepackage{amssymb}
\usepackage{graphicx}
\usepackage[colorlinks,citecolor=blue]{hyperref}
\usepackage{color}
\usepackage{iopams}

\usepackage[title, titletoc]{appendix}

\newcommand{\pf}{{\mathrm{Pf}}}

\begin{document}

\title
{Gutzwiller Approach for Elementary Excitations in $S=1$ Antiferromagnetic Chains}

\author{Zheng-Xin Liu}
\address{Institute for Advanced Study, Tsinghua University, Beijing, 100084, P. R. China}
\author{Yi Zhou}
\address{Department of Physics, Zhejiang University, Hangzhou, 310027, P.R. China}
\author{Tai-Kai Ng}
\address{Department of Physics, Hong Kong University of Science and Technology, Clear Water Bay Road, Kowloon, Hong Kong, China}

\begin{abstract}
In a previous paper [\href{http://prb.aps.org/abstract/PRB/v85/i19/e195144}
{Phys. Rev. B 85,195144 (2012)}], variational Monte Carlo method (based on Gutzwiller projected states) was generalized to $S=1$ systems. This method provided very good trial ground states for the gapped phases of $S=1$ bilinear-biquadratic (BLBQ) Heisenberg chain. In the present paper, we extend the approach to study the low-lying elementary excitations in $S=1$ chains. We calculate the one-magnon and two-magnon excitation spectra of the BLBQ Heisenberg chain and the results agree very well with recent data in literature. In our approach, the difference of the excitation spectrum between the Haldane phase and the dimer phase (such as the even/odd size effect) can be understood from their different topology of corresponding mean field theory. We especially study the Takhtajan-Babujian critical point. Despite the fact that the `elementary excitations' are spin-1 magnons which are different from the spin-1/2 spinons in Bethe solution, we show that the excitation spectrum, critical exponent ($\eta=0.74$) and central charge ($c=1.45$) calculated from our theory agree well with Bethe ansatz solution and conformal field theory predictions. 
\end{abstract}




\tableofcontents

\section{Introduction}

The Haldane phase\cite{Haldane} reveals important physics in $S=1$ spin chains and has been profoundly studied in literature. The Haldane phase has a disordered ground state and a finite excitation gap. Especially, there is spin-$1/2$ edge state at each open boundary\cite{White93,Ng94}. These features can be simply understood in a valence-bond-solid picture proposed by Affleck-Kennedy-Lieb-Tasaki (AKLT)\cite{AKLT}. It was also discovered that the AKLT model and generally any state in the Haldane phase has a hidden $Z_2\times Z_2$ symmetry breaking\cite{Z2Z2break} and a nonzero string order\cite{string}. These nontrivial properties shows that the Haldane phase is distinguished from a trivial phase (such as the dimer phase or the large $D$ phase, where $DS_z^2$ is the single-ion anisotropy term) and was though to be a topological phase. Recently, it was shown that the Haldane phase is protected by symmetry, such as $Z_2\times Z_2$ spin rotation symmetry or time reversal symmetry, and is called a symmetry protected topological (SPT) phase\cite{SPT}. 1-dimensional SPT phases are classified by projective representations of the symmetry group\cite{CGW1DSPT} . New SPT phases as generalizations of the Haldane phase are realized in spin chains or ladders\cite{NewSPT}.

Numerous theoretical methods had been applied to study the Haldane phase, such as effective field theory (via nonlinear sigma model plus a topological theta term)\cite{Haldane}, Bosonization theory\cite{Bosonization}, Schwinger-Boson mean field theory\cite{SchwigerBoson}, fermionic mean field theory\cite{LZN,Cenke Xu}, and various numerical techniques such as density matrix renormalization (DMRG)\cite{White93}, exact diagonalization\cite{ED2003}, and time-evolution-block-decimation(TEBD)\cite{TEBD}. Recently, variational Monte Carlo (VMC) method was generalized to $S=1$ systems\cite{LZTNW} and was applied to study the Haldane phase and the dimer phase of the $S=1$ BLBQ Heisenberg chain. Although the energy obtained is not as accurate as DMRG and TEBD, the advantage of VMC is that we can easily read out the topological structure of ground states in different phases.


The $S=1$ BLBQ Heisenberg model\cite{new,Kato97,BLBQex,BLBQ}
\begin{eqnarray}\label{H}
H=\sum_i[J\mathbf S_i\cdot\mathbf S_{i+1} + K (\mathbf S_i\cdot\mathbf S_{i+1})^2], \ \ (J>0).
\end{eqnarray}
is a variation of the $S=1$ AKLT model. It has attracted much interest in the quantum magnetism community because of its rich phase diagram. 
In the antiferromagnetic section (where we can set $J=1$), the model contains three phases: 
the dimer phase at $K<-1$, the Haldane phase with $-1<K<1$ and a gapless phase at $K>1$. 
In Ref.~\cite{LZTNW}, we revisited this model via VMC method by using Gutzwiller projected $p$-wave Bardeen-Cooper-Schrieffer (BCS) wave functions as trial ground state wavefunctions. We found that the optimized projected BCS wavefunctions are very close to the true ground states for model (\ref{H}) in the region $K \leq1$.  In particular, the optimized projected BCS state is the exact ground state at the AKLT point $K ={1\over3}$. Since the pairing symmetry is $p$-wave, the unprojected BCS states are classified into weak pairing (topologically non-trivial) and strong pairing (topologically trivial) states by their different winding numbers\cite{LZN}. The topology of the BCS state is found to be important in distinguishing Haldane and dimer phases: after Gutzwiller projection the weak pairing states become the Haldane phase whereas the strong pairing states become the dimer phase. The phase transition between the Haldane phase to the dimer phase is reflected as a topological phase transition between weak pairing and strong pairing phases.

Since Gutzwiller projected BCS wavefunction is a resonating valence bound (RVB) state\cite{Anderson87}, or a spin liquid state, our VMC approach reveals that the two gapped phases are two different classes of (fermioinc) RVB states. The topology of the $S=1$ BCS mean field state reflects the pairing pattern of the resulting $S=1$ RVB state: the projected weak/strong pairing state is a long/short ranged RVB state. Here long range means that the pairing amplitude $a_{ij}$ [see eq.~(\ref{aij})] between two spins is finite even if $|i-j|\to\infty$, while short range means that $a_{ij}$ exponentially decays to zero with increasing of distance $|i-j|$. It can be shown straightforwardly that the Haldane phase is long-ranged fermionic RVB states (this is a new interpretation of the Haldane phase) while the dimer phase is short-ranged fermionic RVB states. The transition point, \textit{i.e.} the Takhtajan-Babujian(TB) point, between them is a quasi-long-ranged fermionic RVB state where $a_{ij}$ decays to zero in power law of $|i-j|$.

The success of the Gutzwiller approach in describing the ground states of the BLBQ model leads us to ask the question
that how good this approach is in describing the excited states.
This question is addressed in the present paper. We shall show that the one- and two-magnon excitation spectra calculated numerically from the Gutzwiller projected wavefunctions are consistent with the best available numerical results for the corresponding excitations in the Haldane phase. Interestingly, the excitations in the dimer phase have a very different character - there exists only odd/even-magnon excitations if the length of the chain is odd/even. The TB phase transition point\cite{Takhtajan-1982} between the Haldane and the dimer phases is studied carefully in this paper where we find that the excitation spectrum at the TB point is gapless with the critical exponent and the central charge agree well with $SU(2)_2$ Wess-Zumino-Witten field theory predictions\cite{AffleckWZW}.

This paper is organized as follows. In Section \ref{sec: RevMF}, we review the fermionic mean field theory for $S=1$ model, and discuss the general properties of the corresponding Gutzwiller projected BCS states. The Gutzwiller projected excited states are studied numerically using Monte Carlo technique and the results are presented in section \ref{sec: Num}. Our findings are summarized in section \ref{sec: con} where some general comments to our approach are given.

\section{Fermionic mean-field theory and Gutzwiller Projected states for spin $S=1$ models}\label{sec: RevMF}

Our theory is based on the fermionic representation for $S=1$ systems\cite{LZN,LZTNW}. We introduce three species of fermionic spinons $c_1, c_0, c_{-1}$ to represent the $S=1$ spin operators as $\hat S^a=C^\dag I^a C$, where $a=x,y,z$, $C=(c_1,c_0,c_{-1})^T$ and $I^a$ is the 3 by 3 matrix representation of spin operator. The fermion Hilbert space is identical to the spin Hilbert space when a local particle number constraint $c_1^\dag c_1 + c_0^\dag c_0 +c_{-1}^\dag c_{-1}=1$ is imposed on the system.

In this fermionic representation, the BLBQ model\ (\ref{H}) can be rewritten as
\[
H=-\sum_{\langle i,j\rangle}[J\hat\chi_{ij}^\dag\hat\chi_{ij}+(J-K)\hat\Delta_{ij}^\dag\hat\Delta_{ij}],
\]
where $\hat \chi_{ij}=\sum_{m=1,0,-1}c_{mi}^\dag c_{mj}$ is the fermion hopping operator and $\hat \Delta_{ij}=c_{1i} c_{-1j}-c_{0i}c_{0j}+c_{-1i}c_{1j}$ is the spin-singlet pairing operator. This Hamiltonian can be decoupled in a mean field theory \cite{LZN} by introducing short ranged order parameters $\chi=\langle \hat\chi_{ij}\rangle$, $\Delta=\langle \hat\Delta_{ij}\rangle$, and the Lagrangian multiplier $\lambda$ for the particle number constraint. The mean field Hamiltonian is given by
\begin{eqnarray}\label{Hmf_k}
H_{\mathrm{MF}}&=&\sum_{k}\left[\sum_m\chi_kc_{m,k}^\dag c_{m,k}
-[\Delta_{k}(c_{1,k}^\dag c_{-1,-k}^\dag -{1\over2}c_{0,k}^\dag c_{0,-k}^\dag)+\mathrm{h.c.}]\right]
\nonumber\\
&=&\sum_{m,k\geq0} \varepsilon_k\gamma_{m,k}^\dag\gamma_{m,k}+\textrm{const},
\end{eqnarray}
in momentum space where $\chi_k=\lambda-2J\chi\cos k$, $\Delta_k=-2i(J-K)\Delta\sin k$, and $\varepsilon_k=\sqrt{\chi_k^2+|\Delta_k|^2}$ and $\gamma_{m,k}$ are Bogoliubov eigen-particles [see Eq.~(\ref{Boglbv}) for details]. The mean field Hamiltonian describes a $p$-wave superconductor and may have nontrivial topology. The topology of the mean field states can be more easily seen in Cartesian bases $c_x=(c_{-1}-c_1)/\sqrt2,\ c_y=i(c_{-1}+c_1)/\sqrt2,\ c_z=c_0$, where the mean field Hamiltonian is rewriten as
\begin{eqnarray}
H_{\mathrm{MF}}&=&\sum_{m=x,y,z}\sum_{k}\left[\chi_kc_{m,k}^\dag c_{m,k}+({1\over2}\Delta_{k}c_{m,k}^\dag c_{m,-k}^\dag+\mathrm{h.c.})\right]\nonumber\\
&=& \sum_{m,k} \left(\begin{array}{cc} c_{m,k}^\dag & c_{m,-k}\end{array}\right) \mathcal H_k
\left(\begin{array}{c} c_{m,k} \\c_{m,-k}^\dag\end{array}\right),
\end{eqnarray}
where $\mathcal H_k = {1\over2}(\chi_k\sigma_z + \Delta_k \sigma_y) = {1\over2}\varepsilon_k \pmb\sigma\cdot\pmb n_k$.  Since the unit vector $\pmb n_k$ falls in a circle in $yz$ plain, it defines a map from the momentum space $k$ (a circle) to another circle. 
In analog to the Chern number in 2D, above map has a winding number, 
\begin{eqnarray}\label{winding1}
N_{\rm winding}={1\over2\pi}\int_{-\pi}^{\pi} \hat x\cdot (\pmb n_k\times \partial_k\pmb n_k) dk.
\end{eqnarray}
If $\Delta\neq0$, the topology of the mean-field ground state $|G\rangle_{\mathrm{MF}}$ is determined by $\chi_k$.  If $|\lambda|<|2J\chi|$ (see Fig. \ref{fig:Weak}), the state has winding number 1 for each species of fermions and is called a weak pairing state (i.e. a topological superconductor). On the other hand, if $|\lambda|>|2J\chi|$ (see Fig. \ref{fig:Strong}), the state has winding number 0 and is called a strong pairing state (i.e. a trivial superconductor).

The mean field Hamiltonian (\ref{Hmf_k}) has a global $Z_2$ symmetry, so the mean field state has conserved fermion parity. Furthermore, since mean field parameters are fluctuating, the fermions are effectively coupling to a $Z_2$ gauge field. 
Particularly, in 1D the only effect of the spacial component of the $Z_2$ gauge field is the global $Z_2$ flux, namely, the fermion boundary conditions. We will discuss about the relation between fermion parity and boundary conditions in more detail later.

\subsection{Gutzwiller Projected Ground states}

The mean field ground state $|G\rangle_{\mathrm{MF}}$ is a BCS type wavefunction. After Gutzwiller projection, the state $|\psi\rangle=P_G|G\rangle_{\mathrm{MF}}$ provides a trial ground state for the Hamiltonian (\ref{H}) (see Appendix \ref{append: Gutz} for details). The parameters $\chi,\Delta,\lambda$ are determined by minimizing the energy of the projected states $E_{\mathrm{trial}}=\langle\psi|H|\psi\rangle/\langle\psi|\psi\rangle$ (the details of the calculations can be found in Ref.~\cite{LZTNW}). It was found that the projected weak pairing states corresponds to the Haldane phase and the projected strong pairing states corresponds to the dimerized phase.

A special property of the Gutzwiller projection for spin-1 systems has to be emphasized here. As mentioned above, $S=1$ fermionic mean field states are $p$-wave superconductors, so there are two different topological sectors. 
The fermion boundary conditions have different consequences in different topological sectors. The main issue is the fermion parity. Since the total number of fermions in the system is equal to the number of lattice sites by construction, therefore the fermion parity of the ground state is even/odd for chains with even/odd number of sites. Only the mean field states with proper fermion parity can survive after Gutzwiller projection. 

It was pointed out in Ref.~\cite{Kitaev-2001, LZTNW} that for a weak pairing state, the fermion parity depends on the boundary condition: it is even/odd under anti-periodic/periodic boundary condition. Roughly speaking, this effect is an analogy of 2D band insulators with nonzero winding number (i.e. the Chern number $C$), where a $2\pi$ flux causes fermion number changing by $C$ owning to Hall effect. In our case the flux is quantized in unite of $\pi$ because of pairing. If the winding number is nonzero, then a global $\pi$ flux (which switches the boundary condition) will cause fermion parity change. More precisely, this effect can be easily understood from the dispersion $\chi_k$, as shown in Fig.~\ref{fig:Weak}. We firstly consider the case $L=$even. Under anti-periodic boundary condition, since $\Delta_k$ is always nonzero, the fermions $c_{m,k}, c_{m,-k}$ are paired into Cooper pairs, so the fermion parity is even. Under periodic boundary condition, $\Delta_k$ vanishes at $k=0$ and $k=\pi$, so the fermion modes $c_{m,k=0}$ and $c_{m,k=\pi}$ (where $m=1,0,-1$) are unpaired. Since the chemical potential $|\lambda|<|2J\chi|$, the three fermion modes $c_{m,k=0}$ have negative energy and the other three $c_{m,k=\pi}$ have positive energy. The three modes $c_{m,k=0}$ are occupied in the ground state and therefore the fermion parity is odd. The same results can be obtained for the case $L=$odd using similar arguments.
Therefore when the length of the chain $L$=even/odd, only the anti-periodic/periodic boundary condition survives after Gutzwiller projection in the weak-pairing phase. As a result, the ground state of a closed chain in the Haldane phase is unique. 

In contrast, in the strong pairing phase, the fermion parity is independent on the fermion boundary conditions. We assume $L=$even first. Under anti-periodic boundary condition, the fermion parity is obviously even. Under periodic boundary conditions, the unpaired fermion modes with $k=0$ and $k=\pi$ are unoccupied since they have positive energies, consequently the fermion parity is also even.  In other words, mean field states with both boundary conditions survive after Gutzwiller projection and they have the same energy in thermodynamic limit. So the ground state of the dimer phase is doubly degenerate\cite{LZTNW}. If $L$=odd, the mean field ground state also have even fermion parity under both boundary conditions, but they vanish after Gutzwiller projection. The true ground state of the dimmer model is constructed by Gutzwiller projection of mean field state with one fermion excitation. In that case the ground state is not a spin-singlet.

\textit{Since anti-periodic boundary condition is equivalent to a global $Z_2$ flux through the ring formed by the spin chain, we will denote the ground state with anti-periodic boundary condition as $|\pi$-flux$\rangle$ and denote the one with periodic boundary condition as $|0$-flux$\rangle$ in the following.} The subtle boundary condition effect also exists for the excited states and leads to important distinction between the excitation spectrums in the Haldane and dimer phases as we shall see in the following.

\subsection{Gutzwiller Projected excited states}

In BCS superconductors, excitations are formed by adding quasi-particles obtained from the Bogoliubov-de Gennes equations to the BCS ground  state wavefunction. We may add arbitrary number of quasi-particles to form excited states since the system allows arbitrary fermion number.  We shall assume in the following that (low energy) excitations in the $S=1$ spin liquids can be formed by Gutzwiller projecting the  excited states of the corresponding BCS superconductor. The fixing of fermion parity in spin systems imposes a constraint on the excited states  that can be constructed in this approach.

In the fermionic mean field theory of spin-$1/2$ systems (where the paring symmetry is $s$-wave), the requirement of fixed fermion parity implies that excited states can be formed only by Gutzwiller Projecting BCS excited states with {\em even} number of quasi-particle excitations. The situation is similar for spin-$1$ mean field theory (where the pairing symmetry is $p$-wave) in the strong pairing phase. However, the weak pairing phase is more subtle since the fermion parity can be changed by changing the boundary condition of the mean field Hamiltonian. A consequence is that
one-magnon\cite{note_magnon} excitations are allowed in the Haldane phase.

\subsubsection{Weak pairing phase}

Let us focus on the weak pairing (Haldane) phase. First we consider a spin excitation formed by simultaneously switching the boundary condition and adding a Bogoliubov quasi-particle to the system. We shall call the excitation a one-magnon excitation\cite{note_magnon}. The one-magnon creation operator with $S_z=m$ and momentum $k$ can be written as $\gamma_{m,p}^\dag\hat W$, where $p=k-\pi$ and $\hat W$ is the boundary-twisting operator which switches the periodic boundary condition to anti-periodic and vice versa. Notice that the quasi-particle momentum changes by $\pi$ after the boundary condition is switched (see Appendix \ref{append: momentum} for details). We shall see in next section that after projection the state $P_G\gamma_{m,p}^\dag\hat W|G\rangle_{\mathrm{MF}}$ corresponds to the one-magnon excitation discussed in the literature\cite{White93,TEBD}. Notice that the mean-field energy of this excitation has minimum at $p=0$. This explains why the minimal magnon gap opens at $k=\pi$.  The two-magnon excitation can be obtained by acting the one-magnon creation operator on the mean field ground state twice before the Gutzwiller projection. Notice the boundary condition is restored ($\hat{W}^2=I$) for two-magnons.

In the following we shall provide more details of the one- and two-magnon excitations.

\begin{figure}
\centering
\includegraphics[width=4.5in]{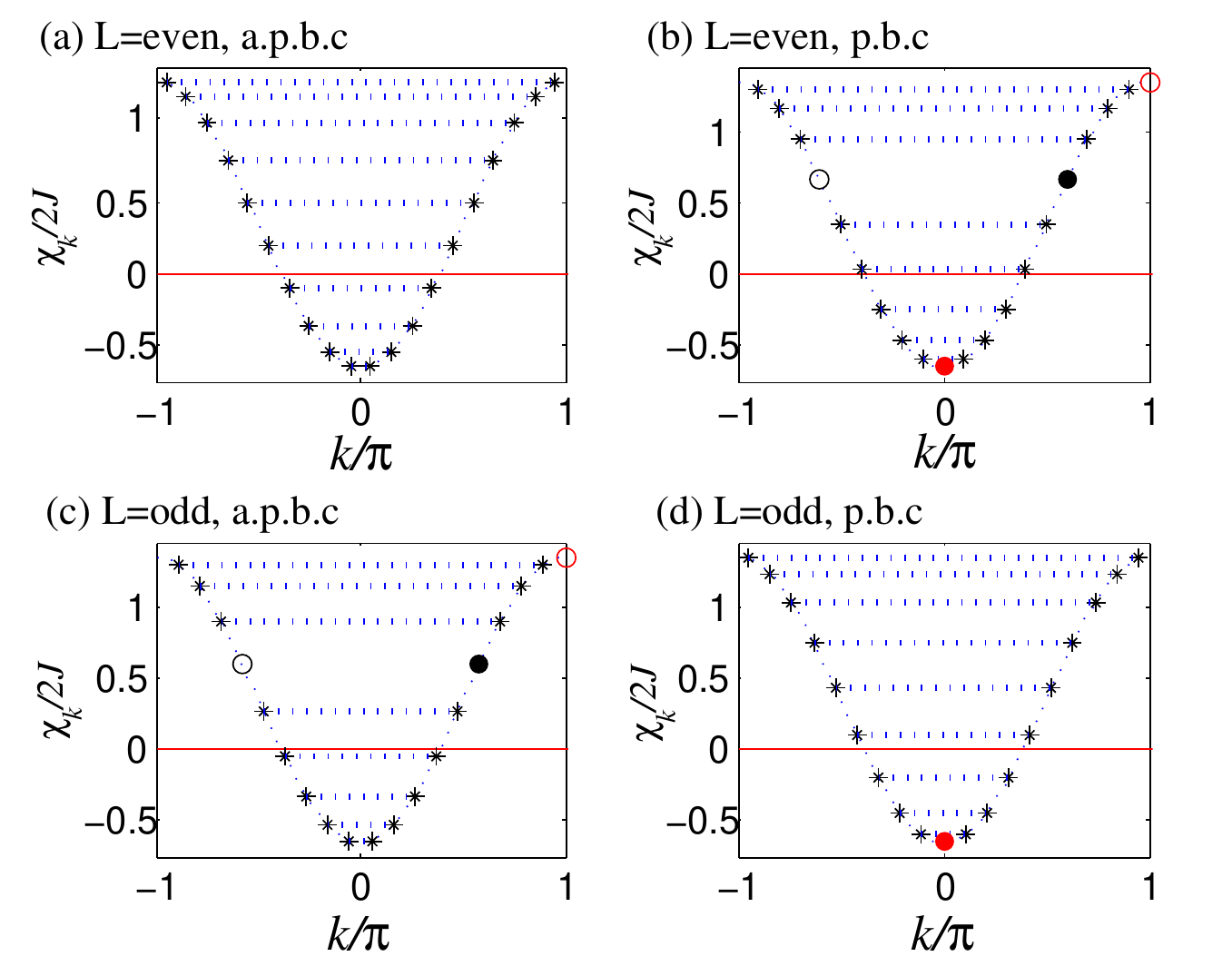}
\caption{(Color online) Dispersion of $\chi_k$ in the weak pairing phase. The red lines show the chemical potential. The dispersion will open a gap if we turn on the paring term $\Delta_k$. The asterisks linked by doted lines show the Cooper pair of spinons. The dots at $k=0$ and $k=\pi$ are marked in red color, meaning that the pairing $\Delta_k$ vanishes at these points. The black solid/hollow dots represent occupied/unoccupied unpaired spinons coming from a broken Cooper pair.
(a) $L=$even, anti-periodic boundary condition(a.p.b.c), no broken Cooper pairs (\textit{ground state}); (b) $L=$even, periodic boundary condition(p.b.c), one broken Cooper pair (\textit{one-magnon excited state}); (c) $L=$odd, a.p.b.c,  one broken Cooper pair (\textit{one-magnon excited state}); (d) $L=$odd, p.b.c,  no broken Cooper pairs (\textit{ground state}). } \label{fig:Weak}
\end{figure}

We first consider the case $L$=even integer. In this case the ground state is a spin-singlet given by [see Fig.\ref{fig:Weak}(a)]
\[
|\textrm{ground}\rangle=P_G|\pi\textrm{-flux}\rangle.
\]
A single magnon is a spin-1 excitation represented by [see Fig.\ref{fig:Weak}(b)]
\[
|(1,m);p+\pi\rangle=P_G\gamma_{m,p}^\dag\hat W|\pi\textrm{-flux}\rangle=P_G\gamma_{m,p}^+|0\textrm{-flux}\rangle,
\]
where $|(1,m);p+\pi\rangle$ indicates that the one-magnon carries spin quantum numbers $(S,m)=(1,m)$ and lattice momentum $p+\pi$. The one-magnon state $|(1,m);p+\pi\rangle$ is orthogonal to the ground state $|\textrm{ground}\rangle$ because it carries both nonzero spin and momentum. The energy-momentum dispersion of the one-magnon spectrum will be discussed in next section.

The two-magnon excitations can be constructed similarly and are denoted by $|(S,m);p,q\rangle$, where $(S,m)$ are the spin quantum numbers and $p, q$ are the momenta carried by the two magnons. Notice that since each magnon carries spin-1, the total spin of two magnons can be $S= 0,1$ or $2$. For example, the states with $S=0,1,2$  and $m=0$ are given by
\begin{eqnarray*}
&&|(0,0);p,q\rangle=P_G(\gamma_{1,p}^\dag\gamma_{-1,q}^\dag+\gamma_{-1,p}^\dag\gamma_{1,q}^\dag-\gamma_{0,p}^\dag\gamma_{0,q}^\dag)  |\pi\textrm{-flux}\rangle,\\
&&|(1,0);p,q\rangle=P_G(\gamma_{1,p}^\dag\gamma_{-1,q}^\dag-\gamma_{-1,p}^\dag\gamma_{1,q}^\dag)  |\pi\textrm{-flux}\rangle,\\
&&|(2,0);p,q\rangle=P_G(\gamma_{1,p}^\dag\gamma_{-1,q}^\dag+\gamma_{-1,p}^\dag\gamma_{1,q}^\dag+2\gamma_{0,p}^\dag\gamma_{0,q}^\dag)  |\pi\textrm{-flux}\rangle.
\end{eqnarray*}
We have dropped some unimportant normalization constants in writing down the above states. Obviously, the two-magnon states are orthogonal to each other because they carry different spin-quantum numbers. It can be also shown that they are orthogonal to the ground state and the one-magnon states\cite{notation}. For a given momentum $k=p+q$, the total energy $E_k$ depends on the momentum distribution $(p,q)$ of the two magnons and the energy-momentum spectrum of the two-magnon states form continuums.

The $L$=odd integer situation can be constructed similarly as for even chains except that
\[ |0\textrm{-flux}\rangle \Longleftrightarrow|\pi\textrm{-flux}\rangle\]
in writing down the ground and excited state wavefunctions.

\subsubsection{Strong pairing phase}

The fermion parity of the spin chain is independent of boundary conditions in the strong pairing phase. As a result the ground states are doubly degenerate and the excitation spectrums are different for chains with even and odd length $L$'s.

\begin{figure}[t]
\centering
\includegraphics[width=4.5in]{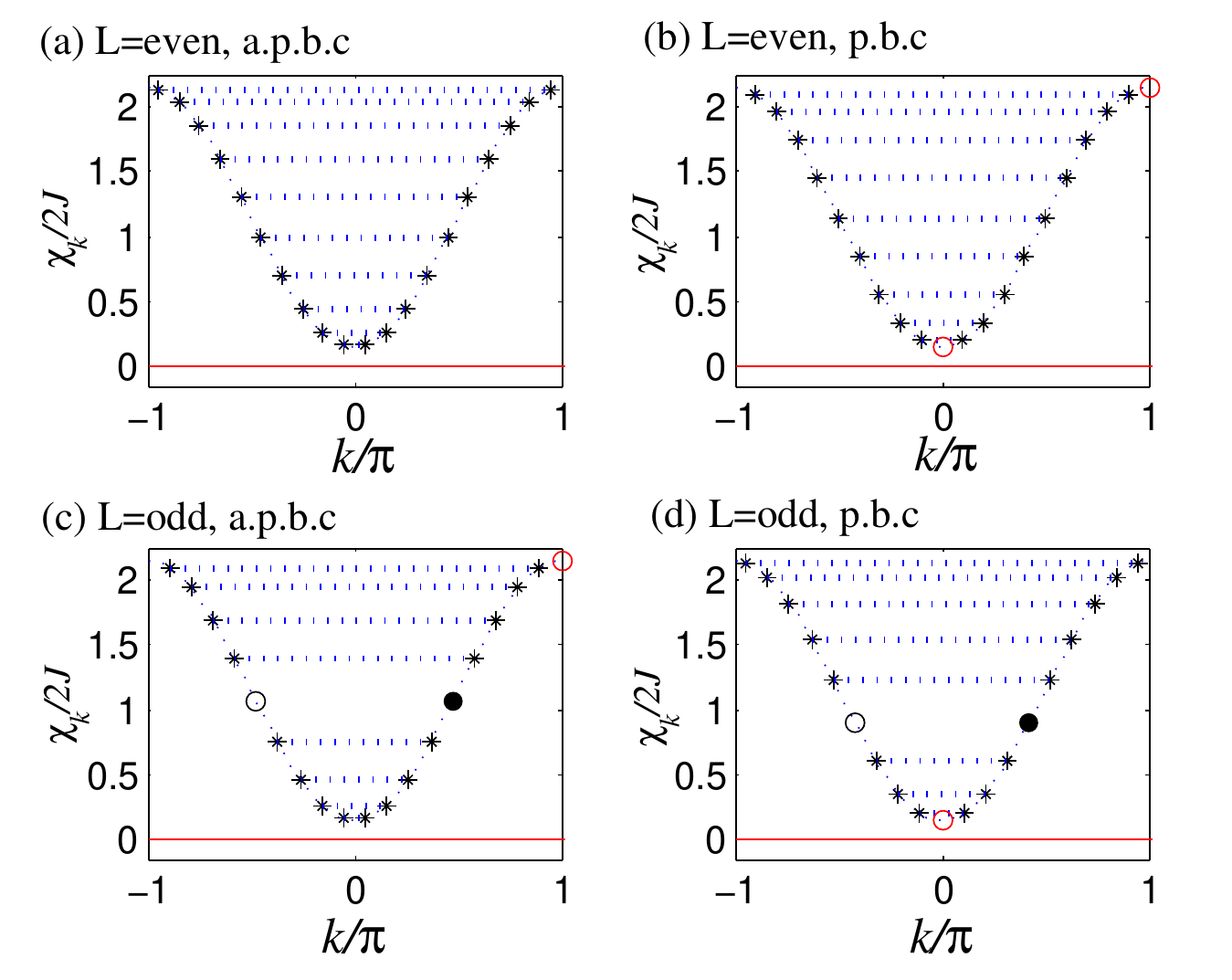}
\caption{(Color online) The dispersion of $\chi_k$ in the strong pairing phase. (a) $L=$even, a.p.b.c, no broken Cooper pairs(\textit{ground state}); (b) $L=$even, p.b.c, no broken Cooper pair (\textit{another ground state}); (c) $L=$odd, a.p.b.c, one broken Cooper pair(\textit{one-magnon excited state}); (d) $L=$odd, p.b.c, one broken Cooper pair(\textit{degenerate one-magnon excited state}). In contrast to the weak pairing phase (see Fig.\ref{fig:Weak}), the spinons at $k=0$ in subfigures (b) and (d) are unoccupied. This is an important difference between the weak pairing phase and the strong pairing phase.
} \label{fig:Strong}
\end{figure}

We consider first the case of $L$=even integer chains. In this case, the ground state wavefunctions are given by [see Fig.\ref{fig:Strong}(a),(b)]
\begin{eqnarray*}
&&|\textrm{ground}\rangle_1=P_G|\pi\textrm{-flux}\rangle,\\
&&|\textrm{ground}\rangle_2=P_G|0\textrm{-flux}\rangle;
\end{eqnarray*}
where $|\textrm{ground}\rangle_1$ carries 0-momentum and $|\textrm{ground}\rangle_2$ carries $\pi$-momentum. One-magnon excitations do not exist in this case since the fermion parity cannot be changed by switching boundary condition. We can only construct two-magnon excitations.

Similar to the ground states, the two-magnon spectra are also doubly degenerate. For simplicity, we only consider excitations above the ground state with $\pi$-flux. Employing the same notation as above, we find that the $|(S,m=0);p,q\rangle$ states are given by
\begin{eqnarray*}
&&|(0,0);p,q\rangle_1=P_G(\gamma_{1,p}^\dag\gamma_{-1,q}^\dag+\gamma_{-1,p}^\dag\gamma_{1,q}^\dag-\gamma_{0,p}^\dag\gamma_{0,q}^\dag)  |\pi\textrm{-flux}\rangle,\\
&&|(1,0);p,q\rangle_1=P_G(\gamma_{1,p}^\dag\gamma_{-1,q}^\dag-\gamma_{-1,p}^\dag\gamma_{1,q}^\dag)  |\pi\textrm{-flux}\rangle,\\
&&|(2,0);p,q\rangle_1=P_G(\gamma_{1,p}^\dag\gamma_{-1,q}^\dag+\gamma_{-1,p}^\dag\gamma_{1,q}^\dag+2\gamma_{0,p}^\dag\gamma_{0,q}^\dag)  |\pi\textrm{-flux}\rangle.\
\end{eqnarray*}

The two magnon excitations form continuum in the energy-momentum spectrum as in the Haldane phase.

Another way to understand why one-magnon excitations do not exist for $L$=even chains in the strong pairing phase is to compare the corresponding mean-field spectra in Fig.\ref{fig:Weak}(b) and Fig.\ref{fig:Strong}(b). We note that the three spinon modes at $k=0$ have negative energy in the weak pairing phase and have positive energy in the strong pairing phase. In the one-magnon excited state of the weak pairing phase (p.b.c), one Bogoliubov quasi-particle is excited whereas the three spinon states at $k=0$ are filled. To construct a similar state in the strong pairing phase, we have to occupy the three spinon states at $k=0$ which corresponds to exciting three (gapped) magnons.  As a result, a one-magnon excited state of the Haldane phase becomes a four-magnon excited state in the dimer phase.

The $L$=odd integer chains have a different character. First of all, the ``ground" state of the system is not a spin singlet but is a spin-triplet with wavefunctions [see Fig.\ref{fig:Strong}(c),(d)]
\begin{eqnarray*}
&&|(1,m);p\rangle_1=P_G\gamma_{m,p}^\dag|0\textrm{-flux}\rangle,\\
&&|(1,m);p+\pi\rangle_2=P_G\gamma_{m,p}^\dag|\pi\textrm{-flux}\rangle;
\end{eqnarray*}
with $m=0,\pm1$, $p=0$ for $|(1,m);p\rangle_1$ and $p=\pi$ for $|(1,m);p+\pi\rangle_2$. The energy of the system changes continuously and forms a one-magnon excitation spectrum when we change $p$. This can be easily understood, since in the dimer phase, the spins form singlet pairs (or dimers) at the ground state. When $L$=odd, not all the spins can form pairs and there must exist odd number of magnons in the system including the ground state.

\section{Numerical results}\label{sec: Num}

In this section we discuss our numerical results for various spin excitations we constructed in the previous section. When $L$ is large, the expectation values of physical quantities in a Gutzwiller projected state can be calculated with Monte Carlo (MC) method\cite{LZTNW,PALee}.

\subsection{Haldane phase: weak pairing state at $K=0$}

\begin{figure}[htbp]
\centering
\includegraphics[width=3.in]{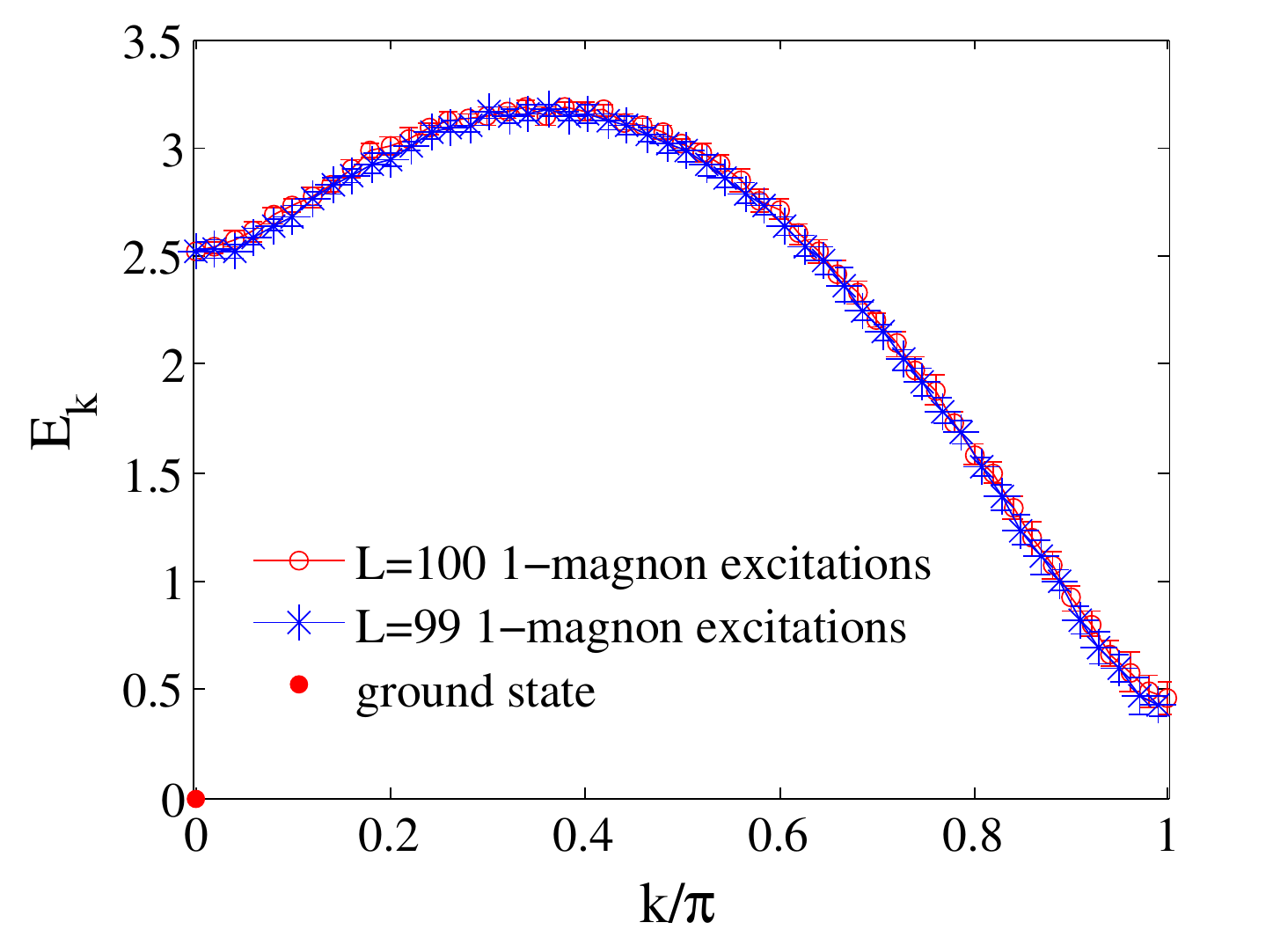}
\caption{(Color online) The dispersion of one-magnon excitations for the Heisenberg model with length $L=$100 and $L=$99. The ground sate energy has been set to 0 and the energy scale is $J=1$. The data for $L$=100 almost coincide with that of $L$=99. The averaged one-magnon gap is $(0.44\pm0.04)J$, which opens at $k=\pi$.
} \label{fig:Heis_Ext}
\end{figure}
\begin{figure}[htbp]
\centering
\includegraphics[width=3.5in]{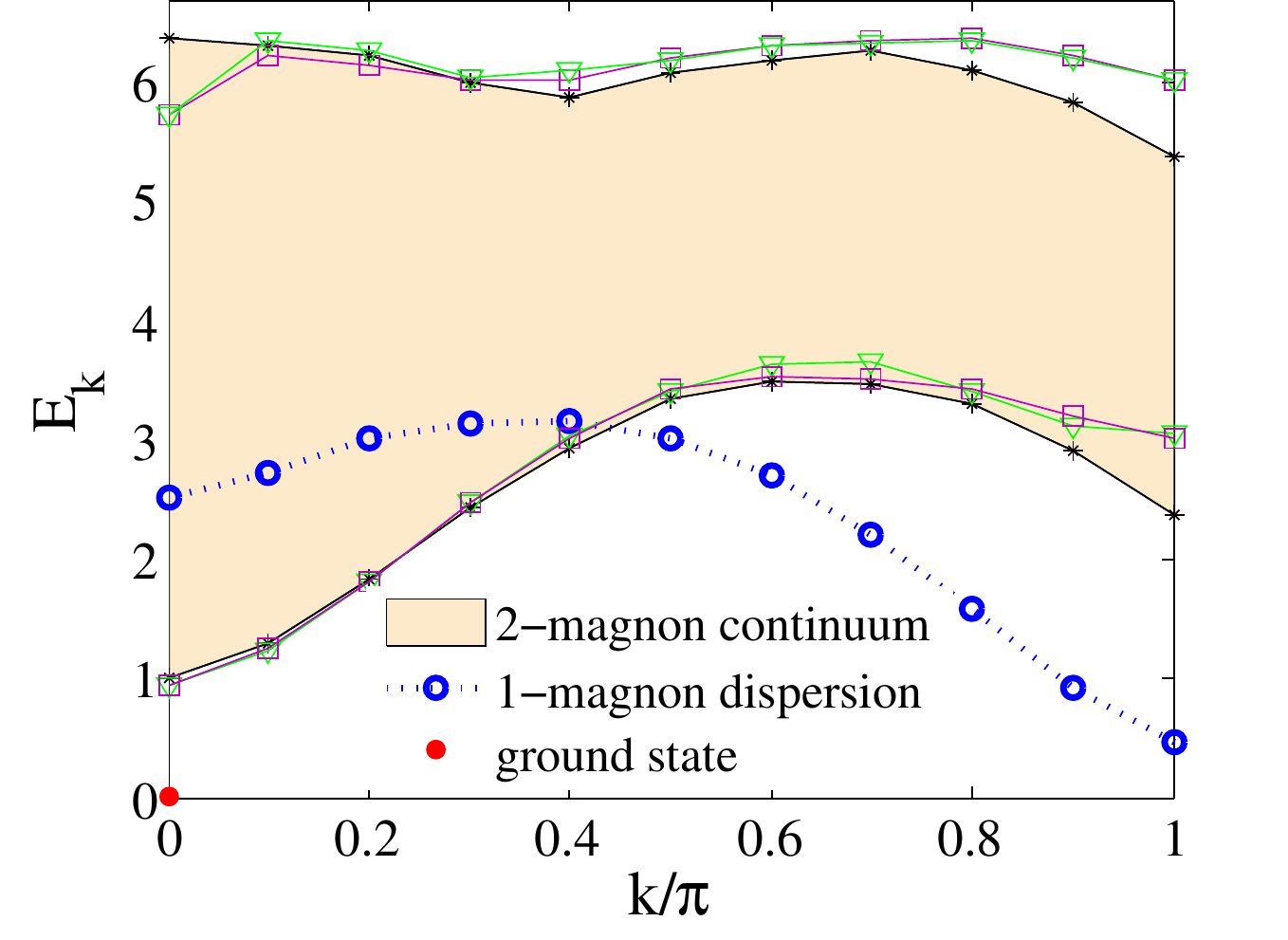}
\caption{(Color online) Low energy excitations of the Heisenberg model ($L=100$). The red solid circle shows the ground state energy. The blue dotted line decorated with hollow circles shows the one-magnon dispersion. The filled area shows the two-magnon-excitation continuum. The two excited magnons can form total spin $S=0,1,2$ states. The black solid lines decorated with asterisks show the upper and lower energy bounds of 2-magnon excitations with total spin-1. Similarly, the purple lines with squares stand for the energy bounds for total spin-2 states, and the green lines with triangles are the energy bounds for total spin-0 states. Similar notations in this figure will be used in Fig.\ref{fig:Dim3_Ext_evenodd} and Fig.\ref{fig:TB_Ext2}.} \label{fig:Heis_Ext_1&2}
\end{figure}

We first consider the Heisenberg model ($K=0$). Fig.~\ref{fig:Heis_Ext} shows the ground state and the one-magnon excitations for two different chains with chain length $L=$100 and $L=$99. We note that the two excitation spectrums almost coincide with each other, showing that even or odd chain length makes little difference in the Haldane phase. The lowest energy one-magnon excitation costs energy $(0.44\pm0.04)J$ and carries momentum $k=\pi$. The maximum of the one-magnon dispersion locates near $k=0.4\pi$. These features agree very well with the numerical results in Ref.~\cite{White93,TEBD} for the one-magnon excitations (where the spin gap is $0.41J$).

The two-magnon excitations form a continuum spectrum, as shown in the filled area in Fig.\ref{fig:Heis_Ext_1&2}. The energy cost for the minimal two-magnon excitation is roughly twice the spin gap. The one-magnon curve merges into the two-magnon continuum below $k=0.4\pi$, suggesting that a single-magnon excitation will decay into two magnons if its momentum is less than $k=0.4\pi$. This result agrees also with the numerical result for the two-magnon spectrum in Ref.~\cite{White93, TEBD}.

Depending on the symmetry under exchanging the spin momentum of the two magnons, the total spin of two magnons can be either 0,2 (symmetric) or 1 (antisymmetric). Fig.\ref{fig:Heis_Ext_1&2} shows that the two-magnon energy bounds for total spin $S=0,1,2$ excitations  are almost the same, with small deviations appearing only near momentum $k=\pi$. This suggests that there is almost no interaction between the two magnons except when their total momentum is close to $k=\pi$. Near $k=\pi$, the $S=1$ channel is lower in energy then the $S=0,2$ channels. Comparing with the energy sum of two one-magnon states, we find that the interaction between the two magnons is attractive for the $S=1$ channel while weakly repulsive for the $S=0,2$ channels, which is qualitatively consistent with Ref.~\cite{White93}. 

\subsection{Dimer phase: Strong pairing state at $K=-3$}

\begin{figure}[htbp]
\centering
\includegraphics[width=6in]{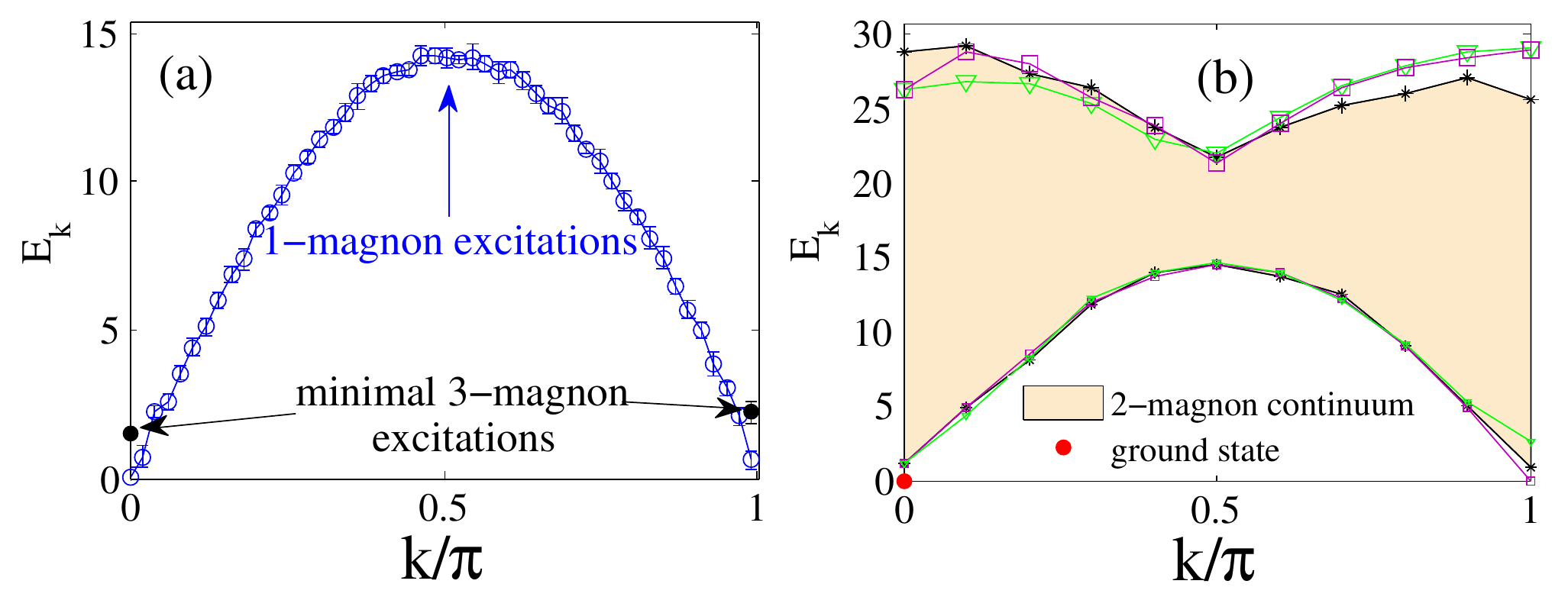}
 \caption{(Color online) The excitations in the strong pairing phase ($K=-3$). (a) $L=99$. The one-magnon excitations. Notice that a singlet (dimer) ground state cannot be constructed for odd $L$. The two solid dark dots show the minimal three-magnon excitation energy at $k=0$ and $k=\pi$ separately; (b) $L=100$. The red solid dot shows the ground state energy and filled area is the  two-magnon continuum.}
\label{fig:Dim3_Ext_evenodd}
\end{figure}

We shall study spin excitations in the strong pairing phase at $K=-3$. As we have pointed out in last section, the $L=$even and $L=$odd chains have quite different properties.  There exist only even/odd-magnon excitations for even/odd $L$.

First we consider even $L$. Fig.~\ref{fig:Dim3_Ext_evenodd}(b) shows the two-magnon continuum for $L=100$. There is an obvious gap (of order $1.1J$) between the ground state and the two-magnon continuum. The two magnons can form states with total spin $S=$0,1 or 2. The energy differences between states with different total spin $S$ are small as is clear from the figure except at the points $k=0,\pi$,  indicating that the two magnons almost do not interact with each other except when their total momentum is close to $k=0$ or $\pi$, similar to the Haldane phase.

Next we consider odd $L$. Recall that the singlet ground state does not exist for odd $L$ and the lowest energy states are the states $|(1,m);p\rangle_1=P_G\gamma_{m,p}^\dag|0\textrm{-flux}\rangle$ with $p=0$, or $|(1,m);p+\pi\rangle_2= P_G\gamma_{m,p}^\dag|\pi\textrm{-flux}\rangle$ with $p=\pi$. The energy dependence of $|(1,m);p\rangle_1$ as function of $p$ is shown in Fig.~\ref{fig:Dim3_Ext_evenodd}(a). We indicate in the figure also the (minimal) 3-magnon excitation energies at points $k=0$ and $k=\pi$. The finite difference in energy between the one- and three- magnon states indicates that the two-magnon excitations have a finite gap of order $1.5J$.

\subsection{TB model: The critical point $K=-1$}
\begin{figure}[htbp]
\centering
\includegraphics[width=6in]{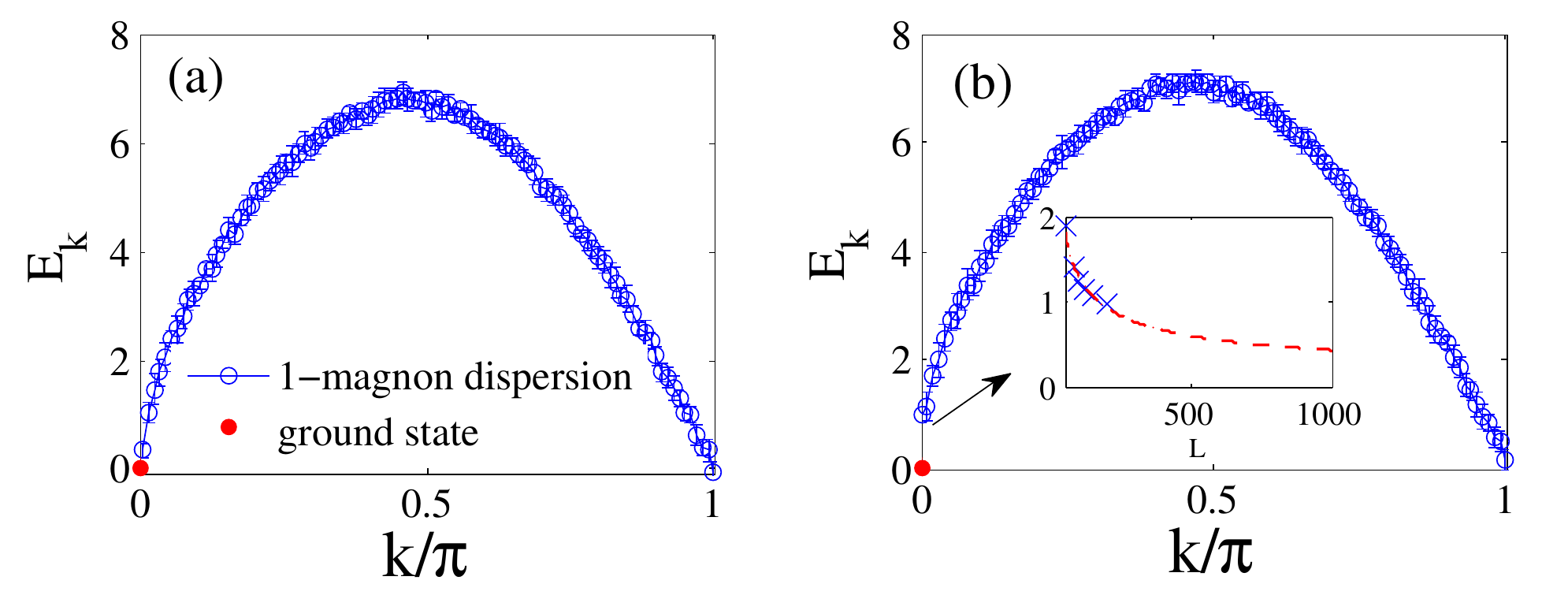}
\caption{(Color online) The one-magnon excitations for the TB point. The results for $L=$even and $L=$odd have a little difference. (a) $L=199$, the one-magnon gap is closing at $k=0$ and $k=\pi$; (b) $L=200$, the one-magnon gap closes at $k=\pi$, while the gap at $k=0$ is finite. The insect shows that the gap at $k=0$ vanishes in power low $L^{-0.48}$. So in
thermodynamic limit the one-magnon excitations  are gapless at $k=0$ and $k=\pi$.
} \label{fig:TB_Ext}
\end{figure}

\begin{figure}[b]
\centering
\includegraphics[width=3.5in]{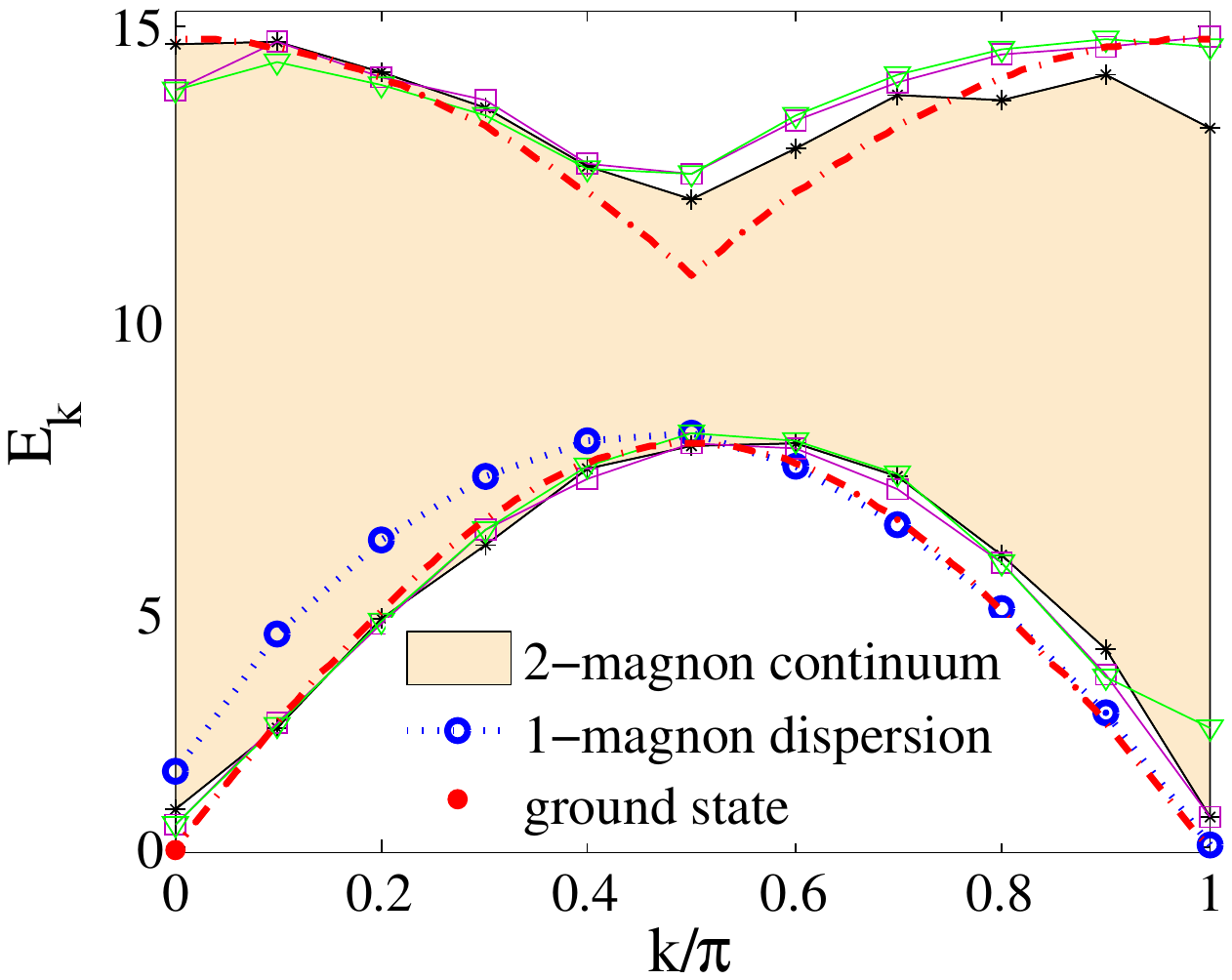}
\caption{(Color online) Excitations at the TB point ($L=100$). The blue dotted line shows the one-magnon dispersion (or the lower bound of four-magnon continuum) and the filled area is the two-magnon continuum of our MC data. The red dash-dotted lines are the boundaries of the two-spinon continuum of the Bethe solution, here we have enlarged the energy scale 1.1 times to fit our data (the inconsistency of energy scales may be caused by finite size effect or systematic error).} \label{fig:TB_Ext2}
\end{figure}

Lastly we consider the TB critical point at $K=-1$ which can be solved exactly with Bethe ansatz\cite{Takhtajan-1982}. At this point, the optimal variational parameters satisfies $\lambda-2\chi\approx 0$ and the mean field excitation spectrum is gapless\cite{LZTNW}. The one-magnon spectrum after Gutzwiller projection is plotted in Fig.~\ref{fig:TB_Ext}. There is a slight difference in energy between $L=$odd and $L=$even chains. Fig.~\ref{fig:TB_Ext}(a) shows that for $L=199$, the spinons at momentum $k=0$ and $k=\pi$ are gapless. Fig.~\ref{fig:TB_Ext}(b) shows the data for $L=200$, the excitation gap closes at $k=\pi$ but remains finite at $k=0$. However, a finite size scaling analysis (insert) shows that the gap at $k=0$ vanishes in power low of the chain length $L$. Thus, we expect that in thermodynamic limit, the one-magnon excitations are gapless at both $k=0$ and $k=\pi$.

The two-magnon continuum for $L=100$ is shown in Fig.\ref{fig:TB_Ext2}. We expect that the one-magnon dispersion will coincide with the lower energy bound of the two-magnon continuum in thermodynamic limit.

\begin{figure}[htbp]
\centering
\includegraphics[width=6in]{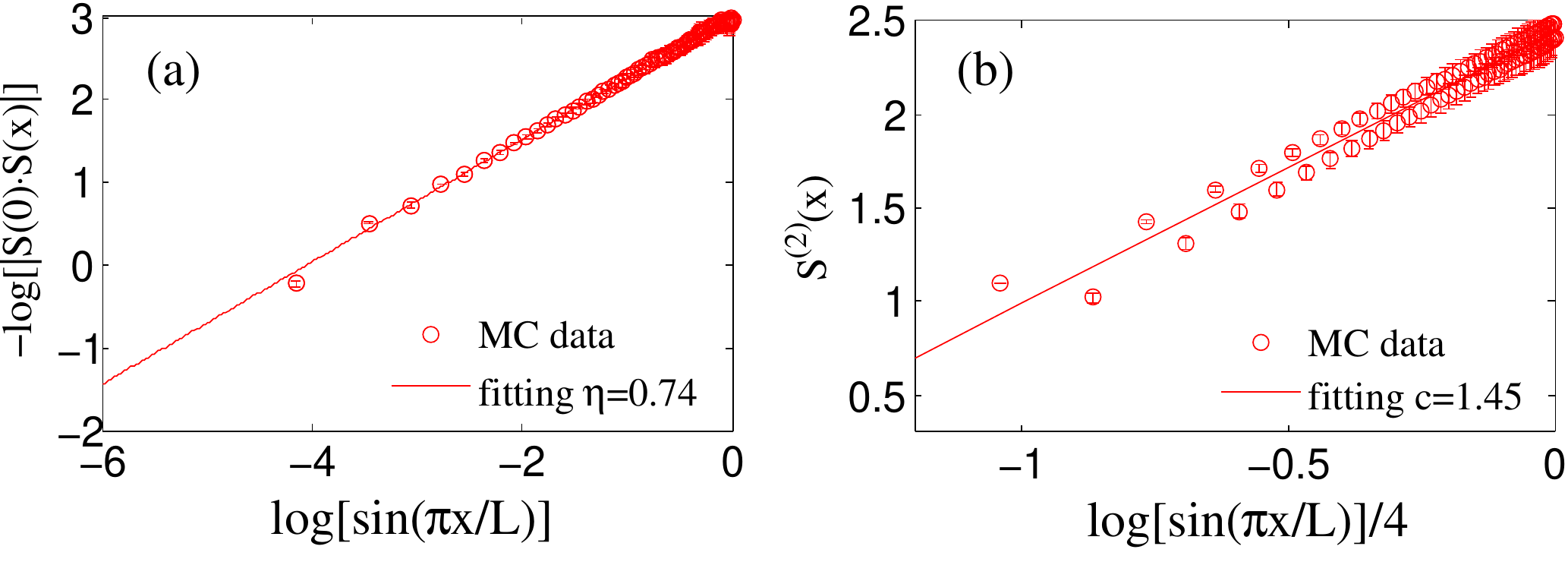}
\caption{(Color online) The critical behavior of the TB model ($L=200$). (a) critical exponent is $\eta=0.74\pm0.01$ fitted by $|\langle\mathbf S_i\cdot \mathbf S_{i+x}\rangle|\propto [\sin({\pi x\over L})]^{-\eta}$; (b) The central charge fitted by $S^{(2)}(x)={c\over4}\log[\sin({\pi x\over L})] + \mathrm{const}$ is $c=1.45\pm0.02$.
} \label{fig:TB_CC}
\end{figure}

We now compare our result with the Bethe ansatz solution\cite{Takhtajan-1982}. In our approach, the elementary excitations are spin-1 magnons whereas the elementary excitations are pairs of spin-1/2 spinons in the Bethe ansatz solution. Therefore, the two approaches do not seem to give the same result at first glance. The correctness of our approach can be verified by checking the critical behavior of the projected state. We numerically calculate the critical exponent $\eta$ and the central charge $c$ from the projected ground state. The results are shown in Fig.\ref{fig:TB_CC}.
The critical exponent is obtained by calculating the spin-spin correlation,
\[
 |\langle\mathbf S_i\cdot \mathbf S_{i+x}\rangle|\propto [\sin({\pi x\over L})]^{-\eta},
\]
 and the central charge is obtained by calculating the second order Renyi entropy\cite{CC},
\[
S^{(2)}(x)={c\over4}\log[{L\over\pi}\sin({\pi x\over L})]+\mathrm{const},
\]
where $S^{(2)}(x)$ is defined as $e^{-S^{(2)}(x)}=\mathrm{Tr}[\rho(x)^2]$ and $\rho(x)$ is the reduced density matrix for a $x$-site subsystem in a $L$-site chain under periodic boundary condition. $\mathrm{Tr}[\rho(x)^2]$ can be calculated with MC technique\cite{CiracSierra,Tarun2013}. We note that $c=0$ for gapped states such as the Haldane phase and the dimer phase, since the Renyi entropy $S^{(2)}(x)$ saturate to a finite constant in large $x$ limit. For the TB model, our results $\eta=0.74\pm0.01,\ c=1.45\pm0.02$ agree very well with $SU(2)_2$ Wess-Zumino-Witten field theory predictions $\eta=0.75,\ c=1.5$\cite{AffleckWZW,Tu}, suggesting that our spectrum is correct in at least the continuum limit.

The agreement of our result with WZW field theory predictions suggests that although the elementary excitations in our approach differ from those in the Bethe ansatz solution, there is a one-to-one mapping between the two approaches in the construction of the  {\em real} spin excitation spectrum. We note that the dispersion of the spin-1/2 spinon in the Bethe Ansatz solution is given by  $\varepsilon(k)={2\pi}\sin|k|$\cite{Takhtajan-1982}, and the excitation spectrum is gapless at $k=0$ and $k=\pi$ in the Bethe-Ansatz solution. The one-magnon dispersion in Fig.~\ref{fig:TB_Ext} is also gapless at $k=0$ and $k=\pi$, and the shape is close to a sine function, in agreement with the Bethe solution. A pair of spin-1/2 spinons form a spin-singlet continuum and a spin-triplet continuum in the Bethe Ansatz solution. The two continuums are degenerate in energy. In our approach, the spin-0 two-magnon continuum and the spin-1 two-magnon continuum are almost degenerate, and correspond to the two continuums of the Bethe solution mentioned above (also see Fig.\ref{fig:TB_Ext2}).

We note also that a one-magnon excited state can also be viewed as a four-magnon excitation in our approach (recall that if one approaches the critical point from the Haldane phase, this state is viewed as a one-magnon state; but if one approaches from the dimer phase, this state is viewed as a four-magnon state), \textit{i.e.} the one-magnon dispersion curve is nothing but the lower bound of the four-magnon continuum and may be constructed from the four- or more-spin-1/2-spinon continuum. Furthermore, the spin-2 two-magnon continuum may correspond to part of the four(or more)-spin-1/2-spinon continuum. These observations suggest that the relation between the $S=1$ magnons in the Gutzwiller projected wavefunction approach and the $S=1/2$ spinons in Bethe Ansatz solution at the TB critical point is highly non-linear\cite{note}.

\section{Conclusion and discussion}\label{sec: con}

\textit{Conclusion} To summarize, we have studied in this paper the low energy spin excitations in the Haldane ($K=0$) and dimer ($K=-3$) phases [including the TB critical point ($K=-1$)] for the one-dimensional BLBQ Heisenberg model using a Gutzwiller Projected wavefunction approach.

We find that the so-called one-magnon excitation observed previously in other numerical methods in the Haldane phase can be explained as a composite object of global $Z_2$ flux and a spinon in our Gutzwiller projected wavefunction approach. The corresponding two-magnon excitation spectrum computed in the Gutzwiller projected wavefunction also agrees with earlier numerical works and we show evidence that the magnons are weakly scattering with each other (absence of confinement).

The excitation spectrum in the dimer phase is computed \textit{(to our knowledge, it is the first time that the energy spectrum of the dimer phase is studied)} where we point out the qualitative differences between $L=$odd and $L=$even chains. At the critical point (the TB model), the projected dispersion is gapless at both $k=0$ and $k=\pi$. The critical exponent $\eta=0.74$ and the central charge $c=1.45$ we obtained agree very well with literature\cite{TEBD, AffleckWZW, Tu}.

\begin{figure}[htbp]
\centering
\includegraphics[width=4.in]{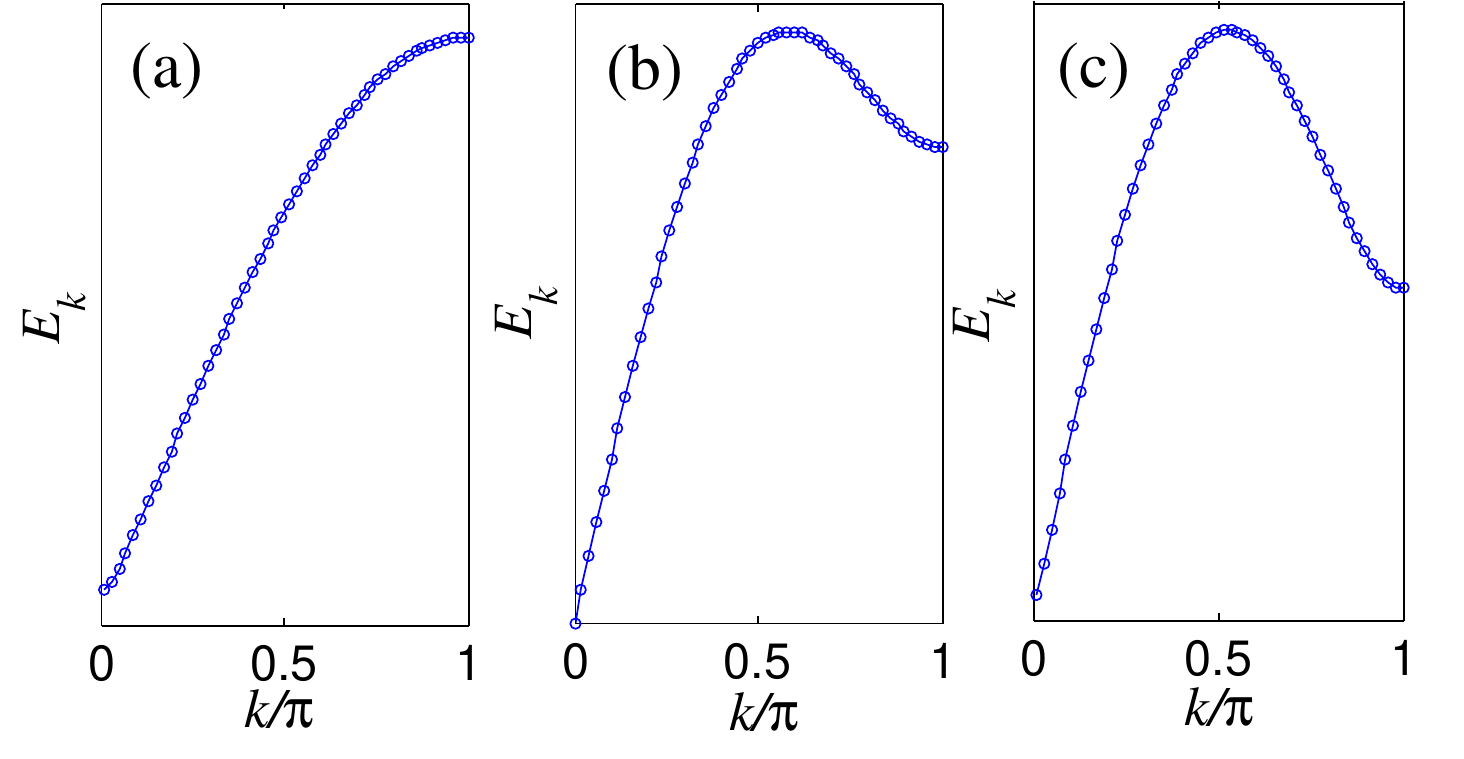}
\caption{(Color online) The mean field dispersion of (a) the weak pairing phase, (b) the critical point, (c) the strong pairing phase. Comparing with Figs.~\ref{fig:Heis_Ext},\ref{fig:TB_Ext},\ref{fig:Dim3_Ext_evenodd}(a), one finds that the one-magnon energy dispersions are dramatically changed after the Gutzwiller projection.
} \label{fig:MF_disp}
\end{figure}

We note that the one-magnon dispersions in Figs.~\ref{fig:Heis_Ext}, \ref{fig:TB_Ext}, \ref{fig:Dim3_Ext_evenodd}(a) are qualitatively different from the corresponding mean field dispersions before Guzwiller projection (see Fig.\ref{fig:MF_disp}). In the weak paring phase, the minimal mean field gap opens at $k=0$, but after projection, the minimal one-magnon gap opens at $k=\pi$. In the strong pairing phase, the mean field dispersion is asymmetric by reflection along $k=0.5\pi$, while after projection the one-magnon curve becomes more symmetric. Especially, the mean field dispersion is gapless only at $k=0$ at the TB point, but the magnons are gapless at both $k=0$ and $k=\pi$ after Gutzwiller projection. These features indicates that only the mean-field states after Gutzwiller projection correctly describe the physical properties of the spin system (\ref{H}).

\textit{Discussion} The existence of one-magnon excitation in the spin-one BLBQ Heisenberg spin chain reflects a fundamental  difference between integer and half-odd-integer spin systems: for integer spin systems, it is possible to form a spin-singlet state for a system with both even and odd number of sites whereas for half-odd-integer spin systems, singlet state exists only in systems with even number of sites. For a spin chain with length $L$, the one-magnon excitation in the Haldane phase can be understood with the single-mode approximation: the ground state (which is a $L$-site singlet state) is reconstructed into a $(L-1)$-site singlet plus a singlet spin. The single spin is propagating and forms as a magnon. It is obvious that it is not possible to form similar excitations above a singlet ground state of half-odd-integer spin systems. The Gutzwiller projected wavefunction approach captures this important difference between integer and half-integer spin systems nicely.

A fundamental question is what's pthe statistics of the magnon excitations. The magnon is a well defined (local) quasi-particle only when the system is gapped.  Since the magnons are created by spin operators, and spin operators at different sites are commuting, two well-separated magnon creation operators commute. So magnons are bosonic. This is consistent with the fact that magnons carry integer spin. In our approach, it seems that the magnons are fermions since we used fermionic spinon representation of spins. However, after Gutzwiller projection, the spinons become local quantities and only the spin configurations and their weight are preserved, the fermionic statistics is discarded. Consequently, the magons(as excited spinons) are bosonic. Statistics of quasi-particles can be changed if the wavefunction obtain a nontrivial Berry phase if one quasi-particle is moved around another one. This situation can take place in 2-dimension, where nonzero flux can be attached to quasi-particles. As a consequence, quasi-particles can obey bosonic, fermionic, anyonic or non-Abelian statistics. Particularly, in our VMC approach, after projection different topology of mean field states may result in different statistics for the quasiparticles.


We thank Hong-Hao Tu for very helpful discussions about the TB point. We also thank T. Senthil, Fan Yang, Fa Wang, Hong Yao, Yao Ma and Cheung Chan for helpful discussions. ZXL is supported by NSFC 11204149 and Tsinghua University Initiative Scientific Research Program. YZ is supported by National Basic Research Program of China (973 Program, No.2011CBA00103), NSFC (No.11074218) and the Fundamental Research Funds for the Central Universities in China. TK Ng acknowledges support by HKRGC grant 603013.

\section*{Appendices}
\begin{appendices}

\section{Details for Gutzwiller projected states}\label{append: Gutz}

\subsection{Bogoliubov eigenstates in mean field theory}\label{Boglbv1}

In momentum space, the mean field Hamiltonian (\ref{Hmf_k}) can be diagnolized into Bogoliubov particles:
\begin{eqnarray}
H&=&\sum_{m,k\geq0}\varepsilon_k\gamma_{m,k}^\dag\gamma_{m,k},\ \ \ \ m=1,0,-1, \label{Boglbv}\\
\gamma_{0,k}&=&u_kc_{0,k} + v_k^* c_{0,-k}^\dag,\nonumber\\
\gamma_{0,-k}^\dag&=&u_kc_{0,-k}^\dag - v_k c_{0,k},\nonumber\\
\gamma_{1,k}&=&u_kc_{1,k} - v_k^* c_{-1,-k}^\dag,\nonumber\\
\gamma_{-1,-k}^\dag&=&u_kc_{-1,-k}^\dag + v_k c_{1,k},\nonumber
\end{eqnarray}
where $\varepsilon_k=\sqrt{\chi_k^2+\Delta_k^2}$, $u_k=\cos{\theta_k\over2}, v_k=i\sin{\theta_k\over2}$ and $\tan\theta_k={i\Delta_k\over\chi_k}$.

In the following, we will provide some eigen states of above Hamiltonian.

(A) Ground state ($E=0$),
\begin{eqnarray}\label{MFBCS}
|\mathrm{G}\rangle_{\mathrm{MF}}&=&\exp\{\sum_{m,n,k}a_{k}\Gamma^{mn}c_{m,k}^\dag c_{n,-k}^\dag\}|\mathrm{vac}\rangle\nonumber\\
&=&\prod_{k}(1+a_{k}\Gamma^{mn}c_{m,k}^\dag c_{n,-k}^\dag)|\mathrm{vac}\rangle\nonumber\\
&=&\prod_{k}(1+a_{k}c_{1,k}^\dag c_{-1,-k}^\dag)\prod_{q>0}(1-a_{q}c_{0,q}^\dag c_{0,-q}^\dag)|\mathrm{vac}\rangle
\end{eqnarray}
where $a_k={v_k^*\over u_k}$ and $\Gamma^{mn}$ is the CG coefficient: $\Gamma^{1,-1}=\Gamma^{-1,1}=-\Gamma^{0,0}=1$ and others equal to zero.

(B) Excited states by breaking the pair $c_{0,p}^\dag c_{0,-p}^\dag$:

1), one-spinon excitation ($E=\varepsilon_p$, \textit{two-fold degenerate})
\begin{eqnarray*}
&&|0;p\rangle_{\mathrm{MF}}=\gamma_{0,p}^\dag |\mathrm{G}\rangle_{\mathrm{MF}}=c_{0,p}^\dag |\mathrm{G}\rangle_{\mathrm{MF}},\\
&&|0;-p\rangle_{\mathrm{MF}}=\gamma_{0,-p}^\dag |\mathrm{G}\rangle_{\mathrm{MF}}=c_{0,-p}^\dag |\mathrm{G}\rangle_{\mathrm{MF}},
\end{eqnarray*}

2), two-spinon excitation ($E=2\varepsilon_p$)
\begin{eqnarray*}
|0,0;p,-p\rangle_{\mathrm{MF}}=(1+a_p^{-1}c_{0,p}^\dag c_{0,-p}^\dag)|\mathrm{G}\rangle'_{\mathrm{MF}}.
\end{eqnarray*}
where $|m;p\rangle_{\mathrm{MF}}$ means the excited spinon carry spin momentum $S_z=m$ and lattice momentum $p$
, and
\[
|\mathrm{G}'\rangle_{\mathrm{MF}}=\prod_{k}(1+a_{k}c_{1,k}^\dag c_{-1,-k}^\dag)\prod_{q>0,q\neq p}(1-a_{q}c_{0,q}^\dag c_{0,-q}^\dag)|\mathrm{vac}\rangle.
\]

(C) Excited states by breaking the pair $c_{1,p}^\dag c_{-1,-p}^\dag$:

1), one-spinon excitation ($E=\varepsilon_p$, \textit{two-fold degenerate})
\begin{eqnarray*}
&&|1;p\rangle_{\mathrm{MF}}=\gamma_{1,p}^\dag |\mathrm{G}\rangle_{\mathrm{MF}}=c_{1,p}^\dag |\mathrm{G}\rangle_{\mathrm{MF}},\\
&&|-1;-p\rangle_{\mathrm{MF}}=\gamma_{-1,-p}^\dag |\mathrm{G}\rangle_{\mathrm{MF}}=c_{-1,-p}^\dag |\mathrm{G}\rangle_{\mathrm{MF}},
\end{eqnarray*}

2), two-spinon excitation ($E=2\varepsilon_p$)
\begin{eqnarray}
|1,-1;p,-p\rangle_{\mathrm{MF}}=(1-a_p^{-1}c_{1,p}^\dag c_{-1,-p}^\dag)|\mathrm{G}\rangle"_{\mathrm{MF}}.
\end{eqnarray}
where
\[
|\mathrm{G}''\rangle_{\mathrm{MF}}=\prod_{k\neq p}(1+a_{k}c_{1,k}^\dag c_{-1,-k}^\dag)\prod_{q>0}(1-a_{q}c_{0,q}^\dag c_{0,-q}^\dag)|\mathrm{vac}\rangle.
\]

Similarly, we can obtain more excited states by breaking more BCS pairs. However, when performing Gutzwiller projection, there will be a subtle problem in the weak pairing phase owning to the dependence of fermion parity on boundary conditions.

\subsection{Projected states in weak pairing phase}

Now we consider the mean field low energy excited states in the weak pairing phase, and their Gutzwiller projection. We will treat $L=$even and $L=$odd separately.

The following property of pfaffian is useful in case that not all fermions are paired. Assuming $A$ is an $n$-dimensional skew symmetric matrix, then we have,
\begin{eqnarray}
\pf A&=&\sum_{i=1}^n(-1)^ia_{1i}\pf A'\nonumber\\
&=&\sum_{i=1}^n(-1)^ia_{ni}\pf A'',
\end{eqnarray}
where $A'$($A''$) mean $A$ with the first($n$th) and the $i$th row and column are removed.

\textbf{ $L$=even}.

1), Ground state

The fermion parity is even under anti-periodic boundary condition (we note it as $|\pi$-flux$\rangle$) and odd under periodic boundary condition ($|0$-flux$\rangle$). Only the former survives after Gutzwiller projection and will be the ground state:
\begin{eqnarray*}
|\mathrm{G}\rangle_{\mathrm{MF}}&=&|\pi\textrm{-flux}\rangle
\nonumber\\
&=&\exp\{\sum_{k}a_{k}c_{1,k}^\dag c_{-1,-k}^\dag\}\exp\{-\sum_{q>0}a_{q}c_{0,q}^\dag c_{0,-q}^\dag\}|\textrm{vac}\rangle\\
&=&\prod_{\{i,j\}}(1+a_{ij}c_{1,i}^\dag c_{-1,j}^\dag)\prod_{\{r,s\}}(1-a_{rs}c_{0,r}^\dag c_{0,s}^\dag)|\textrm{vac}\rangle.
\end{eqnarray*}
where $a_{-k}=-a_k$ and
\begin{eqnarray}\label{aij}
a_{ij}={1\over L}\sum_k a_k\sin[k(i-j)].
\end{eqnarray}

Projected mean field ground state is the approximate ground state: 
\begin{eqnarray}\label{GuzGround}
|\mathrm{ground}\rangle&=&P_G|\textrm{G}\rangle_{\textrm{MF}}\\
&=&\sum_\alpha \textrm{sgn}(i_1,...,i_{n_1},j_1,...,j_{n_1},r_1,...,r_{n_0})
\pf A(\alpha) \pf B(\alpha)|\alpha\rangle\nonumber
\end{eqnarray}
where $i_1,...,i_{n_1}$ ($j_1,...,j_{n_1}$,$r_1,...,r_{n_0}$) are the positions of the 1(-1,0)-component spins in configuration $\alpha$. Obviously, $2n_1+n_0=L$. The sign $\textrm{sgn}(i_1,...,i_{n_1},j_1,...,j_{n_1},r_1,...,r_{n_0})=(-1)^P$, where $P$ is the the permutation number by permuting $i_1,...,i_{n_1},j_1,...,j_{n_1},r_1,...,r_{n_0}$ into the standard order $1,2,...,L$. The matrices $A(\alpha)$ and $B(\alpha)$ are defined as
\begin{eqnarray}
&&A(\alpha)=\left(\begin{array}{cccccc}
&&&a_{i_1j_1}&...&a_{i_1j_{n_1}}\\ &&&\vdots&\ddots&\vdots\\
&&&a_{i_{n_1}j_1}&...&a_{i_{n_1}j_{n_1}}\\a_{j_1i_1}&...&a_{j_1i_{n_1}}&&&\\
\vdots&\ddots&\vdots&&&\\a_{j_{n_1}i_1}&...&a_{j_{n_1}i_{n_1}}&&&\\
\end{array}\right),\label{MatrixA}\\
&&B(\alpha)=\left(\begin{array}{cccc}
0&-a_{r_1r_2}&...&-a_{r_1r_{n_0}}\\
-a_{r_2r_1}&0&...&-a_{r_2r_{n_0}}\\
\vdots&\vdots&\ddots&\vdots\\
-a_{r_{n_0}r_1}&-a_{r_{n_0}r_2}&...&0
\end{array}\right).\label{MatrixB}
\end{eqnarray}

2), 1-magnon excited states

A 1-spinon excited mean field state is given as
\begin{eqnarray*}
|(1,0);p\rangle_{\textrm{MF}}&=&\gamma_{0,p}^\dag \hat W|\pi\textrm{-flux}\rangle=\gamma_{0,p}^\dag |0\textrm{-flux}\rangle\nonumber\\
&=&\prod_{k\neq0}(1+a_{k}c_{1,k}^\dag c_{-1,-k}^\dag)c_{1,0}^\dag c_{-1,0}^\dag
\prod_{q>0}(1-a_{q}c_{0,q}^\dag c_{0,-q}^\dag)c_{0,p}^\dag c_{0,0}^\dag|\textrm{vac}\rangle,
\end{eqnarray*}
where $|(S,m);p\rangle_{\textrm{MF}}$ means that the spinon carries spin-quantum number $(S,m)$ and momentum $p$, and $\hat W$ is the boundary-condition twisting operator that switches periodic boundary condition to anti-periodic boundary condition and vice versa (namely, $\hat W$ adds a global $Z_2$ flux through the whole system).

Gutzwiller projected mean field 1-spinon excited states are approximate 1-magnon excited states:
\begin{eqnarray}\label{PG1mang}
|(1,0);p+\pi\rangle&=&P_G\gamma_{0,p}^\dag \hat W|\pi\textrm{-flux}\rangle\nonumber\\
&=&\sum_\alpha \textrm{sgn}(i_1,...,i_{n_1},j_1,...,j_{n_1},r_1,...,r_{n_0})
\pf A_{00}(\alpha) \pf B_{p0}(\alpha)|\alpha\rangle,
\end{eqnarray}
where
\begin{eqnarray}
&&A_{00}(\alpha)=\left(\begin{array}{cccccccc}
&&&&&& 1& 1\\
&&&&&&\vdots&\vdots\\
&&&&&& 1& 1\\
&&&A(\alpha)&&&-1&1\\
&&&&&&\vdots&\vdots\\
&&&&&&-1&1\\
-1&...&-1& 1&...& 1&0&0\\
-1&...&-1&-1&...&-1&0&0\\
\end{array}\right),\label{MatrixA00}\\
&&B_{p0}(\alpha)=\left(\begin{array}{cccccc}
&&&&e^{ipr_1}&1\\
&&&&e^{ipr_2}&1\\
&&B(\alpha)&&\vdots&\vdots\\
&&&&e^{ipr_{n_0}}&1\\
-e^{ipr_1}&-e^{ipr_2}&...&-e^{ipr_{n_0}}&0&0\\
-1&-1&...&-1&0&0\\
\end{array}\right).\nonumber\label{MatrixBp0}
\end{eqnarray}
Notice that momentum of the magnon is equal to the sum of the spion and the extra $Z_2$ flux, which gives $p+\pi$ (for details, see Section \ref{append: momentum}).

3), 2-magnon excited states

Now we consider two-spinon (or two-magnon) excited states. We can either excite two $c_0$ spinons or one $c_1$ spinon plus one $c_{-1}$ spinon. But these states do not respect the symmetry of the spin Hamiltonian since they do not carry correct spin quantum numbers. According to the total spin ($S$=0,1,2) of the two magnons (the $S$=0,2 states are symmetric under exchanging the spin quantum numbers of the two spinons, while the $S=1$ states are anti-symmetric under exchanging the spin quantum numbers of the two spinons), an excited eigenstate is a superposition the two states listed above. Owning to the degeneracy, we only consider the $(S,0)$-component of the excited states:
\begin{eqnarray}
|(0,0);p,q\rangle_{\textrm{MF}}&=&
(\gamma_{1,p}^\dag\gamma_{-1,q}^\dag+\gamma_{-1,p}^\dag\gamma_{1,q}^\dag-\gamma_{0,p}^\dag\gamma_{0,q}^\dag)
|\textrm{G}\rangle_{\textrm{MF}},\nonumber\\
|(2,0);p,q\rangle_{\textrm{MF}}&=&
(\gamma_{1,p}^\dag\gamma_{-1,q}^\dag+\gamma_{-1,p}^\dag\gamma_{1,q}^\dag+2\gamma_{0,p}^\dag\gamma_{0,q}^\dag)
|\textrm{G}\rangle_{\textrm{MF}},\nonumber\\
|(1,0);p,q\rangle_{\textrm{MF}}&=&
(\gamma_{1,p}^\dag\gamma_{-1,q}^\dag-\gamma_{-1,p}^\dag\gamma_{1,q}^\dag)
|\textrm{G}\rangle_{\textrm{MF}}.
\end{eqnarray}
Above we have assumed that $p+q\neq0$. If $p+q=0$, then the corresponding 2-spinon excited state should be constructed as mentioned in appendix \ref{Boglbv1}.

The corresponding Gutzwiller projected states are listed below:
\begin{eqnarray}
|(0,0);p,q\rangle&=&P_g|(0,0);p,q\rangle_{\textrm{MF}}\nonumber\\
&=&\sum_\alpha \textrm{sgn}(i_1,...,i_{n_1},j_1,...,j_{n_1},r_1,...,r_{n_0}),\nonumber\\
&&\times\left[\pf A^s_{pq}(\alpha) \pf B_{}(\alpha)-\pf A_{}(\alpha) \pf B_{pq}(\alpha)\right]|\alpha\rangle\nonumber\\
|(2,0);p,q\rangle&=&P_g|(2,0);p,q\rangle_{\textrm{MF}}\nonumber\\
&=&\sum_\alpha \textrm{sgn}(i_1,...,i_{n_1},j_1,...,j_{n_1},r_1,...,r_{n_0})\nonumber\\
&&\times\left[\pf A^s_{pq}(\alpha) \pf B_{}(\alpha)+2\pf A_{}(\alpha) \pf B_{pq}(\alpha)\right]|\alpha\rangle,\nonumber\\
|(1,0);p,q\rangle&=&P_g|(1,0);p,q\rangle_{\textrm{MF}}\nonumber\\
&=&\sum_\alpha \textrm{sgn}(i_1,...,i_{n_1},j_1,...,j_{n_1},r_1,...,r_{n_0})\nonumber\\
&&\times\pf A^a_{pq}(\alpha) \pf B_{}(\alpha)|\alpha\rangle.
\end{eqnarray}
where $A(\alpha)$ and $B(\alpha)$ are given in (\ref{MatrixA}) and (\ref{MatrixB}) respectively, and
{
\begin{eqnarray*}
&&A^s_{pq}(\alpha)=\nonumber\\
&&\left(\begin{array}{cccccccc}
&&&&&& e^{ipi_1}& e^{iqi_1}\\
&&&&&&\vdots&\vdots\\
&&&A(\alpha)&&& e^{ipi_{n_1}}& e^{iqi_{n_1}}\\
&&&&&& e^{ipj_{1}}&e^{iqj_{1}}\\
&&&&&&\vdots&\vdots\\
&&&&&& e^{ipj_{n_1}}&e^{iqj_{n_1}}\\
-e^{ipi_1}&...&-e^{ipi_{n_1}}&-e^{ipj_{1}}&...&-e^{ipj_{n_1}}&0&0\\
-e^{iqi_1}&...&-e^{iqi_{n_1}}&-e^{iqj_{1}}&...&-e^{iqj_{n_1}}&0&0\\
\end{array}\right), \\
&&A^a_{pq}(\alpha)=\nonumber\\
&&\left(\begin{array}{cccccccc}
&&&&&& e^{ipi_1}& e^{iqi_1}\\
&&&&&&\vdots&\vdots\\
&&&A(\alpha)&&& e^{ipi_{n_1}}& e^{iqi_{n_1}}\\
&&&&&&-e^{ipj_{1}}&e^{iqj_{1}}\\
&&&&&&\vdots&\vdots\\
&&&&&&-e^{ipj_{n_1}}&e^{iqj_{n_1}}\\
-e^{ipi_1}&...&-e^{ipi_{n_1}}& e^{ipj_{1}}&...& e^{ipj_{n_1}}&0&0\\
-e^{iqi_1}&...&-e^{iqi_{n_1}}&-e^{iqj_{1}}&...&-e^{iqj_{n_1}}&0&0\\
\end{array}\right), \\
\\
&&B_{pq}(\alpha)=\left(\begin{array}{cccccc}
&&&&e^{ipr_1}&e^{iqr_1}\\
&&&&e^{ipr_2}&e^{iqr_2}\\
&&B(\alpha)&&\vdots&\vdots\\
&&&&e^{ipr_{n_0}}&e^{iqr_{n_0}}\\
-e^{ipr_1}&-e^{ipr_2}&...&-e^{ipr_{n_0}}&0&0\\
-e^{iqr_1}&-e^{iqr_2}&...&-e^{iqr_{n_0}}&0&0\\
\end{array}\right).
\end{eqnarray*}
}

{\it \textbf{$L$=odd}.

1), Ground state
\begin{eqnarray}
|\textrm{G}\rangle_{\mathrm{MF}}&=&|0\textrm{-}\mathrm{flux}\rangle\nonumber\\
&=&\exp\{\sum_{k\neq0}a_{k}c_{1,k}^\dag c_{-1,-k}^\dag\}\exp\{\sum_{q>0}a_{q}c_{0,q}^\dag c_{0,-q}^\dag\}
c_{1,0}^\dag c_{-1,0}^\dag c_{0,0}^\dag|\mathrm{vac}\rangle\nonumber\\
&=&\prod_{\{i,j\}}(1+a_{ij}c_{1,i}^\dag c_{-1,j}^\dag)\prod_{\{rs\}}(1-a_{rs}c_{0,r}^\dag c_{0,s}^\dag)\nonumber\\
&&\times(\sum_{i'}c_{1,i'}^\dag) (\sum_{j'}c_{-1,j'}^\dag)(\sum_{r'} c_{0,r'}^\dag)|\mathrm{vac}\rangle,
\end{eqnarray}
where $a_{ij}$ is defined in (\ref{aij}).

The projected mean field ground state is given by
\begin{eqnarray}
P_G|\textrm{G}\rangle_{\textrm{MF}}&=&\sum_\alpha \mathrm{sgn}(i_1,...,i_{n_1},j_1,...,j_{n_1},r_1,...,r_{n_0})\nonumber\\
&&\times\pf A_{00}(\alpha) \pf B_{0}(\alpha)|\alpha\rangle
\end{eqnarray}
where the matrices $A_{00}$ is given in (\ref{MatrixA00}) and $B_0$ is defined as ($n_0$=odd)
\begin{eqnarray*}
&&B_{0}(\alpha)=\left(\begin{array}{ccccc}
0&-a_{r_1r_2}&...&-a_{r_1r_{n_0}}&1\\
-a_{r_2r_1}&0&...&-a_{r_2r_{n_0}}&1\\
\vdots&\vdots&\ddots&\vdots&\vdots\\
-a_{r_{n_0}r_1}&-a_{r_{n_0}r_2}&...&0&1\\
-1&-1&...&-1&0\\
\end{array}\right).\label{MatrixBp}
\end{eqnarray*}

2), 1-magnon excited states
\begin{eqnarray}
|(1,0);p\rangle_{\mathrm{MF}}&=&\gamma_{0,p}^\dag \hat W|0\textrm{-}\mathrm{flux}\rangle=\gamma_{0,p}^\dag |\pi\textrm{-}\mathrm{flux}\rangle\nonumber\\
&=&\prod_{k}(1+a_{k}c_{1,k}^\dag c_{-1,-k}^\dag)
p\prod_{q>0}(1-a_{q}c_{0,q}^\dag c_{0,-q}^\dag)
c_{0,p}^\dag |\mathrm{vac}\rangle, 
\end{eqnarray}

After projection, the excited state is given by
\begin{eqnarray*}
|(1,0);p+\pi\rangle&=&P_G\gamma_{0,p}^\dag \hat W|0\textrm{-}\mathrm{flux}\rangle\\
&=&\sum_\alpha \mathrm{sgn}(i_1,...,i_{n_1},j_1,...,j_{n_1},k_1,...,k_{n_0})
\pf A(\alpha) \pf B_p(\alpha)|\alpha\rangle
\end{eqnarray*}
where the matrices $A(\alpha)$ is given in (\ref{MatrixA}) and $B_p(\alpha)$ is defined as
\begin{eqnarray*}
&&B_{p}(\alpha)=\left(\begin{array}{ccccc}
0&-a_{r_1r_2}&...&-a_{r_1r_{n_0}}&e^{ipr_1}\\
-a_{r_2r_1}&0&...&-a_{r_2r_{n_0}}&e^{ipr_2}\\
\vdots&\vdots&\ddots&\vdots&\vdots\\
-a_{r_{n_0}r_1}&-a_{r_{n_0}r_2}&...&0&e^{ipr_{n_0}}\\
-e^{ipr_1}&-e^{ipr_2}&...&-e^{ipr_{n_0}}&0\\
\end{array}\right).
\end{eqnarray*}

3), 2-magnon excited states

Similar to the 2-spinon excitations for $L=$even, we have
\begin{eqnarray*}
|(0,0);p,q\rangle&=&P_g|(0,0);p,q\rangle_{\mathrm{MF}}\nonumber\\
&=&\sum_\alpha \mathrm{sgn}
\times\left[\pf A^s_{00,pq}(\alpha) \pf B_{0}(\alpha)
-\pf A_{00}(\alpha) \pf B_{0,pq}(\alpha)\right]|\alpha\rangle,\nonumber\\
|(2,0);p,q\rangle&=&P_g|(2,0);p,q\rangle_{\mathrm{MF}}\nonumber\\
&=&\sum_\alpha \mathrm{sgn}
\times\left[\pf A^s_{00,pq}(\alpha) \pf B_{0}(\alpha)
+2\pf A_{00}(\alpha) \pf B_{0,pq}(\alpha)\right]|\alpha\rangle,\nonumber\\
|(1,0);p,q\rangle&=&P_g|(1,0);p,q\rangle_{\mathrm{MF}}\nonumber\\
&=&\sum_\alpha \mathrm{sgn}
\times\pf A^a_{00,pq}(\alpha) \pf B_{0}(\alpha)|\alpha\rangle,
\end{eqnarray*}
where $\mathrm{sgn}=\mathrm{sgn}(i_1,...,i_{n_1},j_1,...,j_{n_1},r_1,...,r_{n_0})$, and the matrices $A$ and $B$ are defined similar to previous cases and will not be repeated here.
}

\subsection{Projected states in the strong pairing phase}

For $L=$even, single-spinon excitations (or generally odd number of spinon excitations) do not exist. And the method to obtain projected two-spinon excited states are similar to the weak pairing phase. When $L=$odd, even-spinon excitations (including `0-spinon excitation' state) are not allowed, and only odd-spinon excitations exist. The method to obtain projected 1-spinon excited states are similar to the weak pairing phase, except that both $P_G\gamma_{0,p}^\dag|\pi$-flux$\rangle$ and $P_G\gamma_{0,p}^\dag|0$-flux$\rangle$ are allowed here.

\section{Momentum of projected ground states and excited states}\label{append: momentum}

Firstly, let us consider the Heisenberg model in the case $L=$even. The ground state is the projected mean field ground state with anti-periodic boundary condition $|\textrm{ground}\rangle=P_G|\pi\textrm{-flux}\rangle$, which yields
\begin{eqnarray}\label{gij_apbc}
a_{i,j+L}=-a_{ij}.
\end{eqnarray}
where $a_{ij}$ is defined in (\ref{aij}), which is antisymmetric  $a_{ij}=-a_{ji}$ and translational invariant $a_{ij}=a(i-j)$.
(\ref{gij_apbc}) is equivalent to $a(r-L)=-a(r)=a(-r)$.

Now we can show the the ground state is translational invariant using above properties. Suppose an arbitrary spin configuration $|\alpha\rangle=|m_1m_2... m_L\rangle$ has a weight [see eq. (\ref{GuzGround})]
\[
f(\alpha)=\mathrm{sgn(\alpha)}\times\pf A(\alpha)\pf B(\alpha).
\]
The weight of translated configuration $T|\alpha\rangle=|m_2m_3... m_Lm_1\rangle$ is
\[
f(T\alpha)=(-1)^{L-1}\mathrm{sgn(\alpha)}\times\pf A(T\alpha)\pf B(T\alpha),
\]
where the phase factor $(-1)^{L-1}$ is owning to moving a fermion from site $1$ to site $L$, and $A(T\alpha), B(T\alpha)$ can be obtained from $A(\alpha), B(\alpha)$ by the following replacement (assuming $i,j\neq L$):
\begin{eqnarray*}
&&a_{ij}\to a_{i+1,j+1}=a_{ij}, \\
&&a_{iL}\to a_{i+1,1}=-a_{iL},\\
&&a_{Lj}\to a_{1,j+1}=-a_{Lj},
\end{eqnarray*}
Thus, $A(T\alpha), B(T\alpha)$ just defer from $A(\alpha), B(\alpha)$ by multiplying a minus sign to the collum $a_{iL}$ and the row $a_{Lj}$. As a result,  we have $\pf A(T\alpha)\pf B(T\alpha)=-\pf A(\alpha)\pf B(\alpha)$ and
\[
f(T\alpha)=(-1)^{L}\mathrm{sgn(\alpha)}\times\pf A(\alpha)\pf B(\alpha)=f(\alpha).
\]
This proves that the projected state has zero lattice momentum.

Now we look at the `one-magnon' excited states. They exist at periodic boundary condition, namely, $a_{i,j+L}=a_{ij}$, or equivalently 
$a(r-L)=a(r)$. Suppose the excited spinon carry momentum $p$. Assuming $p\neq0$, then we have
\[
f(\alpha)=\mathrm{sgn(\alpha)}\times\pf A_{00}(\alpha)\pf B_{p0}(\alpha).
\]
and
\[
f(T\alpha)=(-1)^{L-1}\mathrm{sgn(\alpha)}\times\pf A_{00}(T\alpha)\pf B_{p0}(T\alpha).
\]
$A_{00}(T\alpha), B_{p0}(T\alpha)$ can be obtained from $A_{00}(\alpha), B_{p0}(\alpha)$ by the following replacement (assuming $i,j\neq L$):
\begin{eqnarray*}
&&a_{ij}\to a_{i+1,j+1}=a_{ij}, \\
&&a_{iL}\to a_{i+1,1}=a_{iL},\\
&&a_{Lj}\to a_{1,j+1}=a_{Lj},\\
&&e^{ipr}\to e^{ip(r+1)}.
\end{eqnarray*}
From (\ref{PG1mang}) and (\ref{MatrixA00}), we have $\pf A_{00}(T\alpha)=\pf A_{00}(\alpha),\ \pf B_{p0}(T\alpha)= e^{ip}\pf B_{p0}(\alpha)$. Consequently,
\begin{eqnarray*}
f(T\alpha)&=&(-1)^{L-1}\mathrm{sgn(\alpha)}\times e^{ip}\pf A_{00}(\alpha)\pf B_{p0}(\alpha)\\
&=&e^{i(p+\pi)}f(\alpha).
\end{eqnarray*}
This shows that the total momentum of the wavefunction is $\pi+p$. The lowest energy spinon carry momentum $p=0$, so the lowest-energy `one-magnon' state carry momentum $\pi+p=\pi$ (in other words, the minimal spin gap opens at momentum $k=\pi$).

Repeating above argument, we can show that when $L=$odd, the ground state carry zero momentum, and the lowest-energy `one-magnon' state carry momentum $k=\pi\pm{\pi\over L}$. In thermodynamic limit $L\to\infty$, the minimal one-magnon gap opens at momentum $k=\pi$.

Now we go to the strong pairing phase. When $L=$even, there are two degenerate ground states. Above we have shown that the state $P_G|\pi\textrm{-flux}\rangle$ carries zero momentum. It is easy to show that the other ground state $P_G|0\textrm{-flux}\rangle$ carries $\pi$ momentum. Both states are translationally invariant. However, since they are degenerate, a superposition of these two states is also a ground state of the spin Hamiltonian. The resultant state do not have certain momentum, and is no longer invariant under translation. This is the reason why the ground states have nonzero spin-Peierls correlation.

\end{appendices}



\begin{thebibliography}{10}
\bibitem{note_magnon} It is necessary to clarify our terminology of spinon and magnon in our spin $S=1$ systems. We call spinon (a Gutzwiller projected state of) an unpaired fermion $c_m$ or more generally a Bogoliubov quasi-particle $\gamma_m$ which carries spin-1 (it is quite different from the slave particle approach for spin-1/2 systems where a spinon carries spin-1/2) whereas a magnon is a physical spin-1 excitation in the spin system.  In our Gutzwiller projected wavefunction approach, a magnon in the Haldane phase is a combination of a spinon plus a global $Z_2$ flux (for spin-1/2 systems a magnon is a combined state of two spinons).
\bibitem{note}The fact that different approaches have different `elementary excitations' indicates that for the gapless TB model the `elementary excitations' are not as well-defined as those in gapped phases such as the Haldane phase. Consequently, individual `elementary excitations' in the TB model are not observable. However, the continuous excitation spectrum formed by the `elementary excitations' in different approaches should be the same and can be experimentally detected.
\end{thebibliography}
\end{document}